\documentclass[prd,eqsecnum,twocolumn,amsfonts,amssymb]{revtex4}

\usepackage{graphicx}

\usepackage{bm}

\newcommand{\beq}{\begin{equation}}
\newcommand{\eeq}{\end{equation}}
\newcommand{\beqs}{\begin{eqnarray}}
\newcommand{\eeqs}{\end{eqnarray}}
\newcommand{\lsim}{\mathrel{\raisebox{-
.6ex}{$\stackrel{\textstyle<}{\sim}$}}}

\begin{document}

\title{Higher-Loop Corrections to the Infrared Evolution of a Gauge Theory with
Fermions}

\author{Thomas A. Ryttov}

\author{Robert Shrock}

\affiliation{
C. N. Yang Institute for Theoretical Physics \\
State University of New York \\
Stony Brook, NY 11794}

\begin{abstract}

We consider a vectorial, asymptotically free gauge theory and analyze the
effect of higher-loop corrections to the beta function on the evolution of the
theory from the ultraviolet to the infrared. We study the case in which the
theory contains $N_f$ copies of a fermion transforming according to the
fundamental representation and several higher-dimensional representations of
the gauge group.  We also calculate higher-loop values of the anomalous
dimension of the mass, $\gamma_m$ of $\bar\psi\psi$ at the infrared zero of the
beta function.  We find that for a given theory, the values of $\gamma_m$
calculated to three- and four-loop order, and evaluated at the infrared zero
computed to the same order, tend to be somewhat smaller than the value
calculated to two-loop order.  The results are compared with recent lattice
simulations.

\end{abstract}


\maketitle

\section{Introduction} 

In this paper we investigate how higher-loop corrections to the beta function
affect the evolution of a vectorial, asymptotically free gauge theory (in
$(3+1)$ dimensions, at zero temperature) from the ultraviolet to the infrared.
We assume that the theory contains a certain number, $N_f$, of massless Dirac
fermions $\psi$ transforming according to a representation $R$ of the gauge
group.  We consider cases where $R$ is the fundamental, adjoint, and rank-2
symmetric or antisymmetric tensor representation. We also study the effect of
higher-loop corrections to the anomalous dimension $\gamma_m$ of the fermion
mass.  This work yields more complete information on the nature of the
evolution of the theory from the ultraviolet to the infrared, in particular, on
the determination of the infrared zero of the beta function and the scaling
behavior of the $\bar\psi\psi$ operator in the vicinity of this zero.  We will
give a number of results for a general gauge group $G$ but will focus on the
case $G={\rm SU}(N)$.

We denote the running gauge coupling of the theory as $g(\mu)$, with
$\alpha(\mu)=g(\mu)^2/(4\pi)$, where $\mu$ is the Euclidean energy/momentum
scale (which will often be suppressed in the notation). The property that the
SU($N$) gauge interaction is asymptotically free means that $\lim_{\mu \to
\infty} \alpha(\mu) = 0$, and, since the beta function is negative for small
$\alpha$, it follows that, as the energy/momentum scale $\mu$ decreases from
large values, $\alpha$ increases.  As $\mu$ decreases and the theory evolves
into the infrared, two different types of behavior may occur, depending on the
fermion content.  In a theory with a sufficiently small number of fermions in
small enough representations $R$, as $\mu$ decreases through a scale
$\Lambda$, the coupling $\alpha$ exceeds a critical value $\alpha_{R,cr}$,
depending on $R$, for the formation of bilinear fermion condensates, and these
condensates are produced.  This may or may not be associated with an infrared
(IR) zero of the two-loop beta function at a value $\alpha = \alpha_{IR}$; if
the two-loop ($2\ell$) beta function does have an IR zero, $\alpha_{IR,2\ell}$,
then this type of behavior requires that $\alpha_{IR,2\ell} > \alpha_{R,cr}$
\cite{chipt}, \cite{alm}. As $\mu$ decreases toward $\Lambda$ and $\alpha$
increases toward the IR zero of the beta function, the increase of $\alpha$ as
a function of decreasing $\mu$ is reduced.  This gives rise to an $\alpha$ that
is of order unity, but varies slowly as a function of $\mu$. This behavior
is interestingly different from the behavior of the gauge coupling in quantum
chromodynamics (QCD).  As the condensates form, the fermions gain dynamical
masses of order $\Lambda$, so that in the low-energy effective theory
applicable at scales $\mu < \Lambda$, they are integrated out, and the further
evolution of the theory into the infrared is controlled by the $N_f=0$ beta
function.

Alternatively, if the theory has a sufficiently large number $N_f$ of fermions
and/or if these fermions are in a large enough representation $R$ (as bounded
above by the requirement of asymptotic freedom), then the IR zero of the beta
function occurs at a value smaller than $\alpha_{R,cr}$, so that as the scale
$\mu$ decreases from large values, the theory evolves into the infrared without
ever spontaneously breaking chirally symmetry. In this latter case, the IR zero
of the beta function is an exact IR fixed point (IRFP), approached from below
as $\mu \to 0$.  More complicated behavior occurs in theories
containing fermions in several different representations \cite{bv}; here we
restrict to the case of fermions in a single representation. 
For a given asymptotically free theory that features
an IR fixed point at $\alpha=\alpha_{IR}$, the value of this IRFP decreases as
a function of $N_f$.  There is thus a critical value of $N_f$, denoted
$N_{f,cr}$, depending on $R$, at which $\alpha_{IR}$ decreases below
$\alpha_{R,cr}$.  This value serves as the boundary, as a function of $N_f$,
between the interval of nonzero $1 \le N_f < N_{f,cr}$ where the theory evolves
into the infrared in a manner that involves fermion condensate formation and
associated spontaneous chiral symmetry breaking (S$\chi$SB), and the interval
$N_{f,cr} < N_f < N_{f,max}$, where the theory evolves into the infrared
without this condensate formation and chiral symmetry breaking, with
$N_{f,max}$ denoting the maximal value of $N_f$ consistent with the requirement
of asymptotic freedom.

The anomalous dimension $\gamma_m$ contains important information about the
scaling behavior of the operator $\bar\psi\psi$ for which $m$ is the
coefficient, as probed on different momentum scales. In a theory with an
$\alpha_{IR} \sim O(1)$, it follows that $\gamma_m$ may also be
O(1), which can produce significant enhancement of dynamically generated
fermion masses due to the renormalization-group factor
\beq
\eta = 
\exp \bigg [ \int_{\mu_1}^{\mu_2} \frac{d\mu}{\mu} \gamma_m(\alpha(\mu)) 
\bigg ] \ . 
\label{eta}
\eeq
In a phase where no dynamical mass is generated, $\gamma_m$ simply describes
the scaling behavior of the $\bar\psi\psi$ operator.  

There are several motivations for the study of higher-order corrections to this
evolution of an asymptotically free theory into the infrared.  First, the
critical coupling, $\alpha_{R,cr}$, is generically of order unity, and hence
there is a need to have a quantitative assessment of the importance of
higher-loop corrections to the evolution of the theory.  Second, besides being
of fundamental field-theoretic interest, a knowledge of this evolution plays an
important role in modern technicolor (TC) models with dynamical electroweak
symmetry breaking, in which the slow running of the coupling associated with an
approximate infrared zero of the beta function provide necessary enhancement of
quark and lepton masses \cite{chipt,alm} (recent reviews include
\cite{hillsimmons}-\cite{sanrev}), and can reduce technicolor corrections to
precision electroweak quantities \cite{scalc,ascalc}.  In addition to the
fundamental representation, fermions in higher-dimensional representations have
been studied in the context of technicolor \cite{higherrep,sanrev}.  Fermions
in higher-dimensional representations of chiral gauge groups have long played a
valuable role in studies of extended technicolor (ETC) models that were
reasonably ultraviolet-complete and explicitly worked out the details of the
sequential breaking of the ETC chiral gauge symmetries down to the TC group
\cite{etc}. Recently, there has been a considerable amount of effort devoted to
lattice studies of gauge coupling evolution and condensate formation in
vectorial SU($N$) gauge theories as a function of $N_f$, for fermions in both
the fundamental representation \cite{ascalc},\cite{afn}-\cite{kuti} and higher
representations \cite{cgss}-\cite{cdgk} (a recent review is
\cite{lgtrev}).  Thus, another important motivation for the present work is to
provide higher-order calculations that can be compared with these lattice
studies.

\section{General Theoretical Framework}

\subsection{Beta Function} 

The beta function of the theory is denoted $\beta = dg/dt$, where $dt = d\ln
\mu$. In terms of the variable
\beq
a \equiv \frac{g^2}{16\pi^2} = \frac{\alpha}{4\pi} \ , 
\label{a}
\eeq
the beta function can be written equivalently as $\beta_\alpha \equiv
d\alpha/dt$, expressed as a series 
\beq
\frac{d\alpha}{dt} = -2\alpha \sum_{\ell=1}^\infty b_\ell \, a^\ell  
         = -2\alpha \sum_{\ell=1}^\infty \bar b_\ell \, \alpha^\ell \ , 
\label{beta}
\eeq
where $\ell$ denotes the number of loops involved in the calculation of
$b_\ell$ and $\bar b_\ell = b_\ell/(4\pi)^\ell$.  Although this series and
series for other quantities in quantum field theories do not have finite radii
of convergence but are only asymptotic, experience shows that in situations
where the effective expansion parameter (here, $(\alpha/\pi)$ times appropriate
group invariants) is not too large, the first few terms can provide both
qualitative and quantitative insight into the physics.  The first two
coefficients in the expansion (\ref{beta}), which are scheme-independent, are
\cite{b1}
\beq
b_1 = \frac{1}{3}(11 C_A - 4T_fN_f)
\label{b1}
\eeq
and \cite{b2} 
\beq
b_2=\frac{1}{3}\left [ 34 C_A^2 - 4(5C_A+3C_f)T_f N_f \right ]
\ . 
\label{b2}
\eeq
Here $C_f \equiv C_2(R)$ is the quadratic Casimir invariant for the
representation $R$ to which the $N_f$ fermions belong, $C_A \equiv C_2(G)$ is
the quadratic Casimiar invariant for the adjoint representation, and $T_f
\equiv T(R)$ is the trace invariant for the fermion representation $R$.
Higher-order coefficients, which are scheme-dependent \cite{gw2}, have been
calculated up to four-loop order \cite{b3,b4}. Some further details are given
in Appendix I.

\subsection{Anomalous Dimension of the $\bar\psi\psi$ Operator}

The anomalous dimension $\gamma_m$ for the fermion bilinear $\bar\psi\psi$
describes the scaling properties of this operator and 
can be expressed as a series in $a$ or equivalently, $\alpha$: 
\beq
\gamma_m = \sum_{\ell=1}^\infty c_\ell \, a^\ell 
   = \sum_{\ell=1}^\infty \bar c_\ell \, \alpha^\ell
\label{gamma}
\eeq
where $\bar c_\ell = c_\ell/(4\pi)^\ell$ is the $\ell$-loop series coefficient.
Via Eq. (\ref{eta}), the anomalous dimension $\gamma_m$ governs the running of
a dynamically generated fermion mass.  The coefficients $c_\ell$ have
been calculated to four-loop order \cite{gamma4}.  The first two are
\beq
c_1 = 6 C_f
\label{c1}
\eeq
and
\beq
c_2 = 2 C_f \Big [ \frac{3}{2}C_f + \frac{97}{6}C_A  - \frac{10}{3}
 T_f N_f \Big ] \ . 
\label{c2}
\eeq
For reference, the coefficient $c_3$ is listed in Appendix I.  Since 
as $N_f$ approaches $N_{f,max}$ from below, $b_1 \to 0$ with nonzero $b_2$ 
and hence $\alpha_{IR} \to 0$, and since the perturbative calculations
expresses $\gamma_m$ in a power series in $\alpha$, it follows
that as $\gamma_m \to 0$ as $N_f$ approaches $N_{f,max}$ from
below. We note that a conjectured beta function that directly relates $\beta$ 
to $\gamma$ has been proposed \cite{rs}.

\section{Properties of Beta Function Coefficients and Application to 
Fundamental Representation}

In this section we discuss some general properties of the beta function
coefficients as functions of $N_f$, and give particular results for the case
of fermions in the fundamental representation.  In later sections we consider
fermions in two-index representations.

\subsection{$b_1$}

Since we restrict our considerations to an asymptotically free theory, we
require that, with our sign conventions, $b_1 > 0$.  This, in turn, implies
that
\beq
N_f < N_{f,max} \ , 
\label{nfrange}
\eeq
where
\beq
N_{f,max} = \frac{11C_A}{4T_f} \ . 
\label{nfmax} 
\eeq
Thus, for fermions in the fundamental representation, $N_{f,max,fund}=(11/2)N$.

\subsection{$b_2$ and Condition for Infrared Zero of $\beta$}

We next proceed to characterize the behavior of the higher-loop coefficients of
the beta function, $b_\ell$ with $\ell=2,3,4$, and the resultant zero(s) of the
beta function, in terms of their dependence on $N_f$.  The two-loop results are
well-known and are included here so that the discussion will be self-contained.
Since only the first two coefficients of the beta function are
scheme-independent, it follows that, to the extent that one is in a momentum
regime where one can reliably use the perturbative beta function, the zeros
obtained from these first two coefficients should be sufficient to characterize
the physics at least qualitatively. When one includes higher-loop
contributions to the beta function, one expects shifts of zeros, and there are,
indeed, generically substantial shifts if zeros of the two-loop beta function
occur at $\alpha \sim O(1)$.  However, if inclusion of three- and/or
higher-loop contributions to $\beta$ leads to a qualitative change in behavior,
relative to the behavior obtained from the two-loop $\beta$ function, then the
results cannot be considered fully reliable, since they are scheme-dependent.
For example, for a given gauge group $G$ and fermion content, if the two-loop
beta function does not have an infrared zero but the three-loop beta function
does, one could not conclude reliably that this is a physical prediction of the
theory.  Moreover, it should be noted that even if there is no zero of the
two-loop beta function away from the origin, i.e., a perturbative IRFP, the
beta function may exhibit a nonperturbative slowing of the running associated
with the fact that at energy scales below the confinement scale, the physics is
not accurately described in terms of the Lagrangian degrees of freedom
(fermions and gluons) \cite{btd}-\cite{lmax}.

Another general comment is that the expression of the beta function in
Eq. (\ref{beta}) is semiperturbative and does not incorporate certain
nonperturbative properties of the physics, such as instantons, whose
contributions involve essential zeros of the form ${\rm exp}(-\kappa
\pi/\alpha)$, where $\kappa$ is a numerical constant.  These instanton effects
are absent to any order of the perturbative expansion in Eq. (\ref{beta}) but
play an important role in the theory.  For example, they break the global
U(1)$_A$ symmetry \cite{thooft76} and also enhance spontaneous chiral symmetry
breaking \cite{caldi}-\cite{sv}.  Estimates of the effects of
instantons on the running of $\alpha$ in quantum chromodynamics have found that
they increase this running, i.e., they make $\beta$ more negative in the region
of small to moderate $\alpha$ values \cite{cdg}.  If one were to model
the effect of instantons crudely via a modification of $\beta$ such as 
\beq
\beta_\alpha = \frac{d\alpha}{dt} = - 2\alpha^2 \bigg [ 
\sum_{\ell=1}^\infty \bar b_\ell \, \alpha^{\ell-1} + 
\lambda \, {\rm exp} \Big ( -\frac{\kappa \pi}{\alpha} \Big ) \bigg ] \ , 
\label{betamod}
\eeq
then, since $\lambda > 0$, this would have the effect of increasing
the value of the smallest (nonzero, positive) IR zero $\alpha_{IR}$ of
$\beta$.  For a given minimal value of $\alpha_{cr,R}$ for condensate formation
and spontaneous chiral symmetry breaking, since at least at the perturbative
level $\alpha_{IR}$ is a decreasing function of $N_f$, it would follow that
incorporating instanton effects would increase the value of $N_{f,cr}$, i.e.,
would increase the interval in $N_f$ where there is S$\chi$SB.  Furthermore,
since instantons enhance chiral symmetry breaking, they would tend to reduce
the value of $\alpha_{cr,R}$, which also has the same effect of increasing
$N_{f,cr}$.  We shall comment below on how, although the semiperturbative 
one-gluon exchange approximation to the Dyson-Schwinger (DS) equation does not
directly include effects of confinement or instantons, it may nevertheless
yield an approximately correct value of $N_{f,cr}$ because of another
approximation involved that has the opposite effect on the estimate.  

If one knows the beta function calculated to a maximal loop 
order $\ell_{max}$, then the equation for the zeros of the beta function, 
aside from the zero at $a=0$, is 
\beq 
\sum_{\ell=1}^{\ell_{max}} b_\ell \, a^{\ell-1} = b_1 \bigg [ 1 +
\sum_{\ell=2}^{\ell_{max}} \bigg ( \frac{b_\ell}{b_1} \bigg ) a^{\ell-1} \bigg
] = 0 \ . 
\label{beta_zero}
\eeq
As is clear from Eq. (\ref{beta_zero}), the zeros of $\beta$ away from the
origin depend only on the $\ell_{max}-1$ ratios $b_\ell/b_1$ for $2 \le \ell
\le \ell_{max}$.

The coefficients $b_1$ and $b_2$ are linear functions of $N_f$, while $b_3$ and
$b_4$ are, respectively, quadratic and cubic functions of $N_f$.  With our sign
convention in which an overall minus sign is extracted in Eq. (\ref{beta}),
each of these coefficients is positive for $N_f=0$.  The coefficients $b_1$ and
$b_2$ are both monotonically and linearly decreasing function of $N_f$. 
As $N_f$ increases sufficiently, $b_2$ thus reverses sign, from positive to
negative, vanishing at $N_f=N_{f,b2z}$, where
\beq
N_{f,b2z} = \frac{17 C_A^2}{2T_f(5C_A+3C_f)} \ . 
\label{nfb2z}
\eeq
(The subscript $b \ell z$ stands for the condition that $b_\ell$ is 
\underline{z}ero).  Since
\beqs
N_{f,max} - N_{f,b2z} & = & \frac{3C_A(11C_f+7C_A)}
{4T_f(3C_f + 5C_A)} \cr\cr
                      & > & 0 \ ,
\label{nfdif}
\eeqs
i.e., $N_{f,max} > N_{f,b2z}$, it follows that there is always a nonvacuous
interval in the variable $N_f$ where the theory is asymptotically free and the
two-loop ($2\ell$) beta function has an infrared zero, namely
\beq
N_{f,b2z} < N_f < N_{f,max} \ . 
\label{nfirrange}
\eeq
This zero occurs at
\beq
\alpha_{IR,2\ell} = -\frac{4\pi b_1}{b_2}
\label{alfir_2loop}
\eeq
and is physical for $b_2 < 0$.  Explicitly, for the fundamental representation,
\beq
N_{f,b2z,fund} = \frac{34N^3}{13N^2-3} 
\label{nfb2z_fund}
\eeq
and  
\beq
\alpha_{IR,2\ell,fund} = \frac{4\pi(11N-2N_f)}{-34N^2+N_f(13N-3N^{-1})} \ .
\label{alfir_2loop_fund}
\eeq
The sizes of $\ell$-loop contributions are determined by $(\alpha/\pi)^\ell$
multiplied by corresponding powers of various group invariants.  Illustrative
values of $\alpha_{IR,2\ell,fund}$ are given in Table \ref{betazero} for
$N=2,3,4$ and the subset of the interval (\ref{nfirrange}) for which
$\alpha_{IR,2\ell,fund}$ is not so large as to render the two-loop perturbative
calculation obviously unreliable. Here and below, when $\alpha$ and $\gamma$
values are listed without an explicit $R$, it is understood that they refer to
the fundamental representation. Examples of cases that we do not include in
the table because the two-loop result cannot be considered reliable include the
following (with formal values of $\alpha_{IR,2\ell,fund}$ listed): $N=2$,
$N_f=5$, where $\alpha_{IR,2\ell,fund}=11.4$; $N=3$, $N_f=9$, where
$\alpha_{IR,2\ell,fund}=5.2$; and $N=4$, $N_f=11, \ 12$, where
$\alpha_{IR,2\ell,fund}=14, \ 3.5$. 

For reference, the estimate in Eq. (\ref{alfcrit}) of $\alpha_{cr}$ from the
analysis of the Dyson-Schwinger equation for the fermion propagator, in the
one-gluon exchange approximation, takes the form in Eq. (\ref{alfcrit_fund})
for a fermion in the fundamental representation.  This has the respective
values 1.4, 0.79, and 0.56 for $N=2, \ 3, \ 4$, respectively (where we quote
the results to two significant figures but do not mean to imply that they have
such a high degree of accuracy).  Setting $\alpha_{cr} = \alpha_{IR,2\ell}$
yields the resultant estimates of $N_{f,cr}$, which, rounded to the nearest
integers, are 8, 12, and 16 for these values of $N$.  We denote these as $\beta
DS$ estimates since they combine a calculation of $\alpha_{IR}$ from the
perturbative two-loop $\beta$ function with the (one-gluon exchange
approximation to the) Dyson-Schwinger equation.

As $N_f$ approaches its maximum value, $N_{f,max}$, allowed by the constraint
that the theory be asymptotically free, $b_2$ reaches its most negative value,
namely $b_2 = -C_A(7C_A+11C_f)$.  Clearly, for $N_f$ values such that $b_2$ is
only negative by a small amount and $\alpha_{IR,2\ell}$ is large, the
perturbative calculation is not reliable.  As $N_f$ increases further in the
range (\ref{nfrange}) and $\alpha_{IR,2\ell}$ decreases, the calculation
becomes more reliable.  In Table \ref{nfbz_fund} we list the numerical values
of $N_{f,b2z}$ for some illustrative values of $N$.  At the two-loop level,
depending on whether $\alpha_{IR,2\ell}$ is smaller or larger than a critical
value for fermion condensation, this is an exact or approximate infrared fixed
point (IRFP) of the renormalization group for the gauge coupling.  The
existence of such an IRFP is of fundamental importance in determining how the
theory evolves from the ultraviolet to the infrared \cite{bz}.  In particular,
as mentioned above, this determines whether, as the scale $\mu$ decreases
sufficiently to a scale $\Lambda$ (depending on the group $G$ and the fermion
content), $\alpha$ grows to a large enough size to produce fermion condensates
or, on the contrary, the coupling never gets this large and the theory evolves
into the infrared in a chirally symmetric manner, without ever producing such
fermion condensates.  Note that in the former case, the fermions involved in
the condensates get dynamical masses of order $\Lambda$ and are integrated out
of the effective low-energy field theory applicable for scales $\mu < \Lambda$,
so that the further evolution into the infrared is governed by a different beta
function.

It is useful to observe how rapidly the numbers $N_{f,b2z}$ approach their
large-$N$ values.  The number $N_{f,b2z,fund}$ has the large-$N$ expansion
\begin{widetext}
\beq
N_{f,b2z,fund} = N \bigg [ \frac{34}{13} + \frac{102}{(13N)^2} + 
\frac{306}{(13)^3 N^4} + O\Big (\frac{1}{N^6} \Big ) \bigg ] 
          = N \bigg [ 2.615 + \frac{0.60355}{N^2} + \frac{0.1393}{N^4} 
+ O\Big (\frac{1}{N^6} \Big ) \bigg ] \ . 
\label{nfb2z_fund_taylor}
\eeq
As is evident from Table \ref{nfbz_fund}, the values of $N_{f,b2z,fund}$
approach the leading asymptotic form for moderate $N$, as a result of the fact
that the subleading term in Eq. (\ref{nfb2z_fund_taylor}) is suppressed by 
$1/N^2$. 

It is of interest to consider the 't Hooft large-$N$ limit, where 
\beq
N \to \infty \quad {\rm with} \ \ \alpha N \ \ {\rm fixed }. 
\label{largeNlimit}
\eeq
In a theory with fermions in the fundamental representation, in order for them
to have a non-negligible effect in this limit, one considers the simultaneous
Veneziano limit
\beq
N_f \to \infty \quad {\rm with} \ \  r \equiv \frac{N_f}{N} \ \ {\rm fixed }. 
\label{largeNflimit}
\eeq
In the combined limit of Eqs. (\ref{largeNlimit}) and (\ref{largeNflimit}), 
the range of $r$ satisfying the requirement of asymptotic freedom and the
condition that $b_2 < 0$ so that the two-loop beta function has an IR zero is
\cite{nfintegral} 
\beq
\frac{34}{13} < r < \frac{11}{2} \ , \ i.e., \quad 2.615 < r < 5.5 \ . 
\label{rrange}
\eeq

\subsection{Coefficient $b_3$ and Three-Loop Behavior of the Beta Function}

The three-loop beta function coefficient $b_3$ is a quadratic function of $N_f$
with positive coefficients of its $N_f^0$ and $N_f^2$ terms and a negative
coefficient of its $N_f$ term.  Hence, regarded as a function of the formal
real variable $N_f$, it is positive for large negative and positive $N_f$, and
positive at $N_f=0$.  The derivative of $b_3$ with respect to $N_f$ is
\beq
\frac{db_3}{dN_f} = T_f \bigg [ -\frac{1415}{27}C_A^2 
- \frac{205}{9}C_A C_f + 2C_f^2 + T_f N_f \Big ( \frac{88}{9}C_f
 + \frac{316}{27}C_A \Big ) \bigg ] \ . 
\label{db3dnf}
\eeq
For the fermions representations $R$ that we consider here, for small values of
$N_f$, this derivative $db_3/dN_f$ is negative, so that in this region of
$N_f$, $b_3$ decreases from its positive value at $N_f=0$ as $N_f$ increases.
Because $b_3$ is a quadratic polynomial in $N_f$, the condition that it
vanishes gives two formal solutions for $N_f$, namely
\beq
N_{f,b3z,\pm} = \frac{(1415C_A^2 + 615C_AC_f - 54C_f^2 
\pm 3\sqrt{F_{Rb3}} \ )}{4T_f(79C_A+66C_f )} \quad , \ j=1,2, 
\label{nfb3zpm}
\eeq
and 
\beq
F_{Rb3} = 122157 C_A^4 + 109578C_A^3C_f + 25045 C_A^2 C_f^2
- 7380 C_AC_f^3 + 324 C_f^4 \ . 
\label{frb3}
\eeq
\end{widetext}
Given that $F_{Rb3} > 0$, as is the case here, so that the values $N_{f,b3z,j}$
are real, it follows that $b_3$ is positive in the intervals $N_f <
N_{f,b3z,-}$ and $N_f > N_{f,b3z,+}$ and negative in the interval $N_{f,b3z,-}
< N_f < N_{f,b3z,+}$. The value $N_{f,b3z,+}$ and the neighborhood of $N_f$
values in the vicinity of $N_{f,b3z,+}$ are not of interest here because they
are larger than the maximal value $N_{f,max}$ allowed by the requirement of
asymptotic freedom,
\beq
N_{f,b3z,+} > N_{f,max} \ . 
\label{nfb3z2_ge_nfmax}
\eeq
Thus, $b_3$ only changes sign once for $N_f$ in the asymptotically free
interval $0 \le N_f < N_{f,max}$.  As $N_f$ approaches $N_{f,max}$ from below,
$b_3$ decreases to a negative value given by
\beq
(b_3)_{N_f=N_{f,max}}=-\frac{C_A}{24}\Big [ 1127C_A^2+44C_f(14C_A-3C_f) \Big ]
\ . 
\label{b3nfmax}
\eeq
For fermions in the fundamental representation, this is 
\beq
(b_3)_{N_f=N_{f,max,fund}} = 
-\frac{701}{12}N_c^3+ \frac{121}{12}N_c+ \frac{11}{8N_c}  \ . 
\label{b3nfmax_fund}
\eeq
As is clear from Table \ref{nfbz_fund}, for this case
\beq
N_{f,b3z,1} < N_{f,b2z} \ . 
\label{nfb3z1_le_nfb2z}
\eeq
We noted above that any physically reliable zero of the beta function must be
present already at the level of the two-loop beta function, since this is the
maximal scheme-independent part of this function. Hence, in analyzing such a
zero for the case under consideration where the fermions transform according to
the fundamental representation of SU($N$), we only consider the interval
(\ref{nfirrange}). Combining this fact with our results (\ref{nfb3z1_le_nfb2z})
and (\ref{nfb3z2_ge_nfmax}), it follows that $b_3$ is negative throughout all
of the interval (\ref{nfirrange}) of interest here.  For this
fundamental-representation case, the $N_{f,b3z,j}$ with $j=1,2$ have the
large-$N$ expansions
\beqs
N_{f,b3z,1} & = & N \bigg [ 1.911 + \frac{0.3244}{N^2} + \frac{0.06844}{N^4} 
+ O\Big (\frac{1}{N^6} \Big ) \bigg ] \cr\cr
& & 
\label{nfb3zminus_taylor}
\eeqs
and
\beqs
N_{f,b3z,2} & = & N \bigg [ 13.348 + \frac{1.667}{N^2} + \frac{0.3978}{N^4}
+ O\Big (\frac{1}{N^6} \Big ) \bigg ]  \ . \cr\cr
& 
\label{nfb3zplus_taylor}
\eeqs
Here, $N_{f,max} = 5.5N$.

At three-loop order, the equation for the zeros of the beta function, aside
from the zero at $a=0$, is $b_1 + b_2 a + b_3 a^2=0$.  Formally, this equation
has two solutions for $a$ and hence for $\alpha$, namely
\beq 
\alpha_{\beta z,3\ell,\pm} = 
\frac{2\pi}{b_3}\left [ -b_2 \pm \sqrt{b_2^2-4b_1b_3} \ \right ] \ . 
\label{alf_betazero_3loop}
\eeq
Since $b_2$ must be negative in order for the beta function to have a
scheme-independent infrared zero, and since for fermions in the fundamental
representation we have shown that $b_3 < 0$ in the relevant interval 
(\ref{nfirrange}), we can rewrite this equivalently as 
\beq
\alpha_{\beta z,3\ell,\pm} = \frac{2\pi}{|b_3|}\left [ -|b_2| \mp 
\sqrt{b_2^2+4b_1|b_3|} \ \right ] \ . 
\label{alfir_3loop_manifest}
\eeq

In order for a given solution to be physical, it must be real and positive.  As
is evident from Eq. (\ref{alfir_3loop_manifest}), the solution corresponding to
the the $+$ sign in Eq. (\ref{alf_betazero_3loop}) (i.e., the $-$ sign in
Eq. (\ref{alfir_3loop_manifest}) ) is negative and hence unphysical.  Thus,
there is a unique physical solution for the IR zero of the beta function to
three-loop order, namely 
\beq
\alpha_{IR,3\ell} = \alpha_{\beta z, 3\ell,-} \ .
\label{alfir_3loop}
\eeq
Illustrative values for this IR zero of the beta function at three-loop order
are listed in Table \ref{betazero}. 

For an arbitrary fermion representation for which $\beta$ has a two-loop IR
zero, we observe that the value of this zero decreases when one calculates
it to three-loop order, i.e., 
\beq
\alpha_{IR,3\ell} < \alpha_{IR,2\ell} \ .
\label{alfdif23loops}
\eeq
This can be proved as follows.  We have 
\beq
\alpha_{IR,2\ell} - \alpha_{IR,3\ell} = \frac{2\pi}{|b_2 b_3|}
\bigg [ 2b_1 |b_3| + b_2^2 -|b_2|\sqrt{b_2^2+4b_1|b_3|} \ \bigg ] \ . 
\label{alfdif23loopstart}
\eeq
The expression in square brackets is positive if and only if 
\beq
(2b_1 |b_3|+b_2)^2-b_2^2(b_2^2+4b_1|b_3|) > 0 \ . 
\label{alfdif23loopsaux1}
\eeq
But the difference in (\ref{alfdif23loopsaux1}) is equal to the
positive-definite quantity $b_1^2b_3^2$, which proves the inequality
(\ref{alfdif23loops}).  This inequality is evident in Table \ref{betazero}.

\subsection{Coefficient $b_4$ and Four-Loop Behavior of $\beta$ }

The four-loop beta-function coefficient, $b_4$, was calculated in
Ref. \cite{b4}.  We next analyze its behavior as a function of $N_f$ and the
result four-loop IR zero of the beta function.  The coefficient $b_4$ is a
cubic polynomial in $N_f$ which has positive coefficients of its $N_f^0$, and
$N_f^3$ terms.  Hence, regarded as a function of the formal real variable
$N_f$, $b_4$ is negative for large negative $N_f$, positive for $N_f=0$, and
also positive for large positive $N_f$.  For fermions in the fundamental
representation, the derivative at $N_f=0$ is
\begin{widetext}
\beq
\Bigg ( \frac{db_4}{dN_f} \Bigg )_{N_f=0} = 
-\bigg ( \frac{485513}{1944} + \frac{20}{9}\zeta(3) \bigg ) N^3 
+ \bigg ( \frac{58583}{1944} - \frac{548}{9}\zeta(3) \bigg ) N 
+ \bigg ( -\frac{2341}{216} + \frac{44}{9}\zeta(3) \bigg ) N^{-1} 
- \frac{23}{8}N^{-3} 
\label{db4dnf_nfzero}
\eeq
\end{widetext}
where $\zeta(z)$ is the Riemann zeta function, 
\beq
\zeta(s) = \sum_{n=1}^\infty \frac{1}{n^s}
\label{zeta}
\eeq
and $\zeta(3)=1.20205690...$. This derivative is negative for all
$N$. (In the complex $N$ plane, it has six zeros at three complex-conjugate
pairs of $N$ values.) It follows that, again as a function of the formal
real variable $N_f$, $b_4$ has a local maximum at a negative value of $N_f$ and
then decreases through positive values as $N_f$ increases toward 0 and passes
through 0 into the interval of physical values.  

The detailed behavior of $b_4$ in the physical asymptotically free interval $0
\le N_f \le N_{f,max}$ depends on $N$.  In particular, one may determine the
value of $N_f$ where $b_4$ has a minimum and whether $b_4$ has any zeros for
positive $N_f$.  For SU(2), $b_4$ decreases to a minimum positive value as
$N_f$ ascends through the approximate value $N_f = 5.8$, and then increases
monotonically for larger $N_f$, so that it is positive-definite for all
non-negative $N_f$, in particular, the asymptotically free region $0 \le N_f <
11$.  For SU(3), $b_4$ is also positive-definite for all non-negative $N_f$,
reaching a local minimum as $N_f$ ascends through a value of approximately 8.2
and then increasing monotonically for larger $N_f$. However, for SU(4), $b_4$
is positive for $0 \le N_f \le 9.51$, negative for the interval $9.51 \le N_f
\le 11.83$, and positive again for $N_f > 11.83$, with zeros at $N_f \simeq
9.51$ and $N_f\simeq 11.83$.  We list these zeros of $b_4$ as a function of
$N_f$ in Table \ref{nfbz_fund}. (Again, we recall that the physical values of
$N_f$ are, of course, restricted to non-negative integers.)  Thus, this
reversal of sign occurs in the interval of interest here, $0 \le N_f < 22$,
where the SU(4) theory is asymptotically free.  For SU(5), $b_4$ behaves in a
manner qualitatively similar to the SU(4) case; it is positive for $0 \le N_f
\le 11.18$, negative in the interval $11.18 \le N_f \le 15.18$, and positive
for larger values of $N_f$, vanishing at $N_f \simeq 11.18$ and $N_f \simeq
15.18$.  Thus, again, $b_4$ reverses sign in the region $0 \le N_f < 22.5$
where the SU(5) theory is asymptotically free.  Thus, in contrast with $b_2$
and $b_3$, which are negative throughout the interval of $N_f$ of interest (and
$b_1$, which is positive), $b_4$ can, for $N \ge 4$, vanish and reverse sign in
this interval.

At the four-loop level, the equation for the zeros of the beta function, aside
from $a=0$, is the cubic equation 
\beq
b_1 + b_2 a + b_3 a^2 + b_4 a^3 =0 \ .
\label{beta_zero_4loop}
\eeq
This equation has three solutions for $a$ and hence for $\alpha$, which will be
denoted $\alpha_{\beta z,4\ell,j}$, $j=1,2,3$.  Since the coefficients $b_\ell$
are real, there are two generic possibilities for these three roots, namely
that they are all real, or that one is real and the other two form a
complex-conjugate pair.  The properties of the roots are further restricted by
the asymptotic freedom condition that $b_1 > 0$, the existence of a two-loop IR
zero, which requires that $b_2 < 0$, and the fact that, as we have shown, for
the relevant range (\ref{nfirrange}) of $N_f$, where these conditions are met,
$b_3 < 0$.  As is evident in Table \ref{betazero}, we find that for the values
of $N$ and $N_f$ that we consider, the roots of Eq. (\ref{beta_zero_4loop}) are
real.  For all of the values of $N$ and $N_f$ where there is a reliable
two-loop value for an IR zero of the beta function (i.e., where it does not
occur at such a large value of $\alpha$ as to render the perturbative
calculation untrustworthy), one of these roots is negative and hence not
physical, one of them, namely the minimal positive one, is the physical IR
zero, which we will denote $a_{IR,4\ell}=\alpha_{IR,4\ell}/(4\pi)$, and there
is a third root at a larger positive value.  This third root, denoted
$a_{4\ell,u}=\alpha_{4\ell,u}/(4\pi)$, is not relevant for our analysis, since
the initial value of $\alpha$ at a high energy scale $\mu$ is assumed to be
close to zero, so that as the scale $\mu$ decreases, $\alpha$ increases and
approaches the (positive) zero of the beta function closest to the origin,
namely $\alpha_{IR,4\ell}$ \cite{uvfp}. 

It is straightforward to display the analytic expressions for the root
$\alpha_{IR,4\ell}$, but we shall not need this for our analysis.  We list
numerical values for $\alpha_{IR,4\ell}$ for various values of $N$ and $N_f$ in
Table \ref{betazero}.  For completeness, we note the specific sets $(N,N_f)$
where $\alpha_{IR,2\ell}$ is so large that we consider the analysis via the
perturbative beta function unreliable: these are $(N,N_f)=(2,6)$, (3,9),
(4,11), and (4,12).

\subsection{Estimates of Zeros of the Four-Loop Beta Function via 
Pad\'e Approximants} 

For the beta function, or more
conveniently, the reduced function with the prefactor removed,
$\sum_{j=0}^{\ell_{max}-1} b_j a^{j-1}$, 
it is useful to calculate and analyze Pad\'e approximants, since these provide
closed-form expressions that, by construction, agree with the series to the
maximal order to which it is calculated.  The expansion for $\bar \beta_\alpha$
to $\ell=4$ loop order can be used in two ways.  First, one can simply solve
the cubic equation $\bar\beta_\alpha = b_1 + b_2 a + b_3 a^2 + b_4 a^3 = 0$ and
obtain the three roots, one of which is the root of interest, giving the IR
zero.  Secondly, one can calculate Pad\'e approximants, e.g., the
[2,1] and [1,2] approximants, and determine their zeros. The [1,2]
Pad\'e approximant has a single zero at
\beq
a_{\beta z,4\ell,[1,2]} = \frac{\alpha_{IR,4\ell,[1,2]}}{4\pi} = 
\frac{b_1(b_1b_3-b_2^2)}{b_2^3-2b_1b_2b_3+b_1^2b_4)} \ . 
\label{pade12zero}
\eeq
Taking into account the fact that $b_2$ and $b_3$ are negative in the relevant
interval (\ref{nfirrange}), this can be rewritten as
\beq
a_{\beta z,4\ell,[1,2]} = \frac{\alpha_{IR,4\ell,[1,2]}}{4\pi} = 
\frac{b_1(b_1|b_3|+b_2^2)}{|b_2|^3+2b_1|b_2||b_3|-b_1^2b_4)} \ . 
\label{pade12zero_manifest}
\eeq
The two zeros from the [2,1] approximant are
\beqs
& & a_{\beta z,4\ell,[2,1],\pm} = \cr\cr
& & \frac{ b_2b_3 - b_1b_4 \pm 
\Big [ (b_2b_3 - b_1b_4)^2 - 4b_1b_3(b_3^2-b_2b_4) \Big ]^{1/2}}
{2(b_2b_4 - b_3^2)} \ . 
\cr\cr
& & 
\label{pade21zeros}
\eeqs
Taking account of the fact that $b_2$ and $b_3$ are negative in the
relevant interval (\ref{nfirrange}), this can be rewritten as
\begin{widetext}
\beq
a_{\beta z,4\ell,[2,1],\pm} = 
\frac{ b_1b_4 - |b_2||b_3| \mp
\Big [ (|b_2||b_3| - b_1b_4)^2 + 4b_1|b_3|(b_3^2+|b_2|b_4) \Big ]^{1/2}}
{2(|b_2|b_4 + b_3^2)}  \ . 
\label{pade21zeros_manifest}
\eeq
\end{widetext}
The expression in Eq. (\ref{pade21zeros_manifest}) with the $-$ sign in front
of the square root is negative and unphysical, while the expression with the
$+$ sign in front of the square root yields the estimate of the IR fixed point,
as $\alpha_{IR,4\ell,[2,1]}=4\pi a_{\beta z,4\ell,[2,1]}$. As is evident from
Eqs. (\ref{pade12zero}) and (\ref{pade21zeros}), the zeros of the [1,2] and
[2,1] Pad\'e approximants incorporate information on $\beta$ up to four loops.
One readily verifies that in the limit $b_4 \to 0$, the zero of the [1,2]
Pad\'e reduces to the two-loop result $a=-b_1/b_2$, and the two zeros of the
$[2,1]$ Pad\'e reduce to those obtained from the three-loop beta function,
(\ref{alfir_3loop}).  We list the values of $\alpha_{IR}$ obtained from the
zeros of the $[1,2]$ and $[2,1]$ Pad\'e approximants to the four-loop beta
function for the case of fermions in the fundamental representation in Table
\ref{betazero}. 

From our calculations of $\alpha_{IR}$ at the three- and four-loop level for
SU($N$) with fermions in the fundamental representation, we can make several
remarks.  Although $n$-loop calculations of the beta function for $n \ge 3$
loops are scheme-dependent, the results obtained with the present
$\overline{MS}$ scheme provide a quantitative measure of the accuracy of the
scheme-independent two-loop result.  For a given $N$, as $N_f$ increases above
the minimal value $N_{f,b2z}$, where the IR zero first appears, and as the
resultant $\alpha_{IR,2\ell}$ decreases to values $\lsim 1$, the difference
between $\alpha_{IR,2\ell}$ and the higher-loop values $\alpha_{IR,n\ell}$ for
$n=3,\ 4$ decrease.  As is evident from Table \ref{betazero}, the value of
$\alpha_{IR,n\ell}$ generically decreases as one goes from $n=2$ to $n=3$ loops
and then increases by a smaller amount as one goes from $n=3$ to $n=4$ loops,
so that $\alpha_{IR,4\ell}$ is smaller than $\alpha_{IR,2\ell}$.  In the same
region of $N_f$ values such that $\alpha_{IR,2\ell}$ is reasonably small, the
values obtained via the [1,2] and [2,1] Pad\'e approximants to the four-loop
beta function are close to those obtained from the zeros of this beta function
itself. 

\section{Evaluation of the Anomalous Dimension $\gamma_m$ at the
Infrared Zero of $\beta$ } 

In this section we evaluate the anomalous dimension of $\gamma \equiv
\gamma_m$, calculated to the $n$-loop order in perturbation theory, at the
(approximate or exact) IR zero of the beta function to this order,
$\alpha_{IR,n\ell}$, for $n=2,3,4$.  We denote these as
$\gamma_{n\ell}(\alpha_{IR,n\ell})$.  We focus here on general results and
their application to the case of fermions in the fundamental representation,
and discuss higher-dimensional representations in subsequent sections. In
general, this anomalous dimension must be positive to avoid unphysical
singularities in fermion correlation functions.

A running fermion mass, $\Sigma(k)$, that is dynamically generated
at a scale $\Lambda$, decays with Euclidean momentum $k > \Lambda$ like 
\beq
\Sigma(k) \sim \Lambda \bigg ( \frac{\Lambda}{k} \bigg )^{2-\gamma_m}
\label{sigmak}
\eeq
up to logs.  Since for $k > \Lambda$, the running coupling $\alpha$ is smaller
than the critical value $\alpha_{R,cr}$ and there is no spontaneous chiral
symmetry breaking, it follows that $\Sigma(k)$ must decrease toward zero as
$k/\Lambda \to \infty$.  In turn, this implies that $\gamma_m < 2$.  Hence, a
physical value of $\gamma_m$ must lie in the range
\beq
0 < \gamma_m < 2 \ . 
\label{gammarange}
\eeq
For values of $N_f$ such that the theory evolves into the infrared in a
chirally symmetric manner, so that the IR zero of the beta function is exact, 
the same upper bound follows from a related unitarity consideration
\cite{ungam}.

Using the two-loop result for $\gamma$ and evaluating it at the two-loop value
of the IR zero of the beta function, we have
\begin{widetext}
\beq
\gamma_{2\ell}(\alpha_{IR,2\ell}) = \frac{C_f(11C_A-4T_fN_f)(455C_A^2+99C_AC_f
+(180C_f-248C_A)T_fN_f +80T_f^2N_f^2)}{12(-17C_A^2+(10C_A+6C_f)T_fN_f)^2}
\label{gamma_irfp_2loop}
\eeq
For the fundamental representation, this is
\beq
\gamma_{2\ell}(\alpha_{IR,2\ell}) = \frac{(N^2-1)(11N-2N_f)
(1009N^3-99N-(158N^2+90)N_f+40NN_f^2)}{12(-34N^3+(13N^2-3)N_f)^2}
\label{gamma_irfp_2loop_fund}
\eeq
\end{widetext}
We list numerical values of $\gamma(\alpha_{IR,2\ell})$ in Table
\ref{gammavalues} for the illustrative values $N=2, \ 3, \ 4$ and, for each
$N$, a set of $N_f$ values in the range (\ref{nfirrange}).  For sufficiently
small $N_f > N_{f,b2z}$ in each $N$ case, $\alpha_{IR,2\ell}$ is so large that
the formal value of $\gamma_{2\ell}(\alpha_{IR,2\ell})$ is larger than 2 and
hence unphysical; we enclose these values in parentheses to indicate that they
are unphysical artifacts of a perturbative calculation at an exessively large
value of $\alpha$.

In the large-$N$, large-$N_f$ limit of Eqs. (\ref{largeNlimit}) and
(\ref{largeNflimit}) with $r \equiv N_f/N$, Eq. (\ref{gamma_irfp_2loop_fund})
reduces to
\beq
\gamma_{2\ell}(\alpha_{IR,2\ell}) = \frac{(11-2r)(1009-158r+40r^2)}
{12(-34+13r)^2} + O \bigg ( \frac{1}{N^2} \bigg ) \ . 
\label{gamma_irfp_2loop_fund_largeNNf}
\eeq
For $r=4$ corresponding to the asymptotic value of $N_{f,cr,fund}$ in
Eq. (\ref{nfcr_fund}), $\gamma_{2\ell}(\alpha_{IR,2\ell}) = 113/144 \simeq
0.785$, which is the same as the large-$N$ limit of
Eq. (\ref{gamma_2loop_at_alfcrit_fund_nfcr_taylor}).

One may evaluate $\gamma_{2\ell}(\alpha_{IR,2\ell})$ at $N_f$ equal to the
value $N_{f,cr,fund}$ predicted by the one-gluon exchange (ladder)
approximation to the Dyson-Schwinger equation for the fermion propagator, given
in Eq. (\ref{nfcr_fund}).  This is somewhat formal, since these values of
$N_{f,cr,fund}$ are not, in general, integers and hence not actually physical;
for example, $N_{f,cr,fund} = 7.86, \ 11.91, \ 15.94$ for $N=2, \ 3, \ 4$).
This procedure yields the result
\beq
\gamma_{2\ell}(\alpha_{IR,2\ell};N_{f,cr,fund}) = 
\frac{565N^4-706N^2+225}{144(N^2-1)(5N^2-3)} \ . 
\label{gamma_irfp_2loop_fund_nfcr}
\eeq
For the illustrative cases $N=2, \ 3, \ 4$, this anomalous dimension takes the
values 0.88, 0.82, and 0.80, respectively.  As $N \to \infty$, Eq. 
(\ref{gamma_irfp_2loop_fund_nfcr}) has the expansion 
\beq
\gamma_{2\ell}(\alpha_{cr,fund}) = \frac{113}{144} + \frac{11}{40N^2} + 
O \bigg ( \frac{1}{N^4} \bigg ) \ . 
\label{gamma_2loop_at_alfcrit_fund_nfcr_taylor}
\eeq
Since the estimate (\ref{nfcr_fund}) is close to $4N$ even for the smallest 
value, $N=2$, and asymptotically approaches $4N$ as $N \to \infty$, it is
worthwhile to compare the above values of $\gamma$, viz., 0.88, 0.82, and 0.80
for $N=2, \ 3, \ 4$, with $\gamma_{2\ell}(\alpha_{IR,2\ell})$ evaluated at the
nearest physical, integer values of $N_f$, namely $N_f=8, \ 12, \ 16$ for $N=2,
\ 3, \ 4$.  This procedure yields $\gamma_{2\ell}(\alpha_{IR,2\ell})= 0.75, \
0.77, \ 0.78$, as recorded in Table \ref{gammavalues}.  To within the
strong-coupling theoretical uncertainties of these calculations, these values
are mutually consistent.

A closely related approach is to evaluate the two-loop expression for 
$\gamma_m$ at $\alpha=\alpha_{cr,R}$, where 
$\alpha_{cr,R}$ is the estimate of the critical coupling for fermion
condensation obtained from the one-gluon exchange approximation to the 
Dyson-Schwinger equation, given in Eq. (\ref{alfcrit}).  
If one were to use the first term in 
Eq. (\ref{gamma}) by itself, one would obtain, to this order, 
$\gamma_m = \bar c_1(\alpha_{cr,R}/\pi) = 1/2$, for
an arbitrary fermion representation $R$. The second term, 
$\bar c_2(\alpha_{cr,R}/\pi)^2$, does depend on $R$.  Combining these, i.e.,
evaluating $\bar c_1(\alpha/\pi)+\bar c_2(\alpha/\pi)^2$ at $\alpha = 
\alpha_{cr,R}$, we find, for a general fermion representation $R$, 
\beq
\gamma_{2\ell}(\alpha_{cr,R}) = \frac{97C_A+225C_f-20T_fN_f}{432C_f} \ . 
\label{gamma_2loop_at_alfcrit}
\eeq
For consistency, one would then substitute $N_f$ equal to the estimate 
$N_{f,cr,R}$ from the same Dyson-Schwinger analysis, given in
Eq. (\ref{nfcr}).  For a general representation this yields 
\beq
\gamma_{2\ell}(\alpha_{cr,R};N_f=N_{f,cr,R}) = \frac{21C_A^2+128C_AC_f+
225C_f^2}{144C_f(C_A+3C_f)} \ . 
\label{gamma_2loop_at_alfcrit_nfcr}
\eeq
For the fundamental representation, this reduces to the same result as was 
obtained in Eq. (\ref{gamma_irfp_2loop_fund_nfcr}). 
We have also evaluated the three-loop result for $\gamma$ at the three-loop
value of the IR zero of the beta function, which we denote as
$\gamma_{3\ell}(\alpha_{IR,3\ell})$, and the four-loop result for $\gamma$ at
the four-loop value of the IRFP, which we denote as
$\gamma_{4\ell}(\alpha_{IR,4\ell})$.  We list the resultant values in Table
\ref{gammavalues}. 

From our calculations of $\gamma_m$ for the case of fermions in the fundamental
representation, we can make several observations.  Although computations of
$\alpha_{IR,n\ell}$ and $\gamma_{n\ell}(\alpha_{IR,n\ell})$ are
scheme-dependent for $n \ge 3$ loops, they provide a useful measure of the
accuracy of the lowest-order results.  As was the case with the position of
$\alpha_{IR,n\ell}$ itself, we find that, for a given $N$ and for $N_f$
reasonably well above $N_{f,b2z}$ so that the perturbative calculation of
$\alpha_{IR,n\ell}$ is not too large, the value of
$\gamma_{n\ell}(\alpha_{IR,n\ell})$ generically decreases as one goes from
$n=2$ to $n=3$ loops.  Some of this decrease can be ascribed to the decrease in
$\alpha_{IR,n\ell}$ going from $(n=2)$-loop to $(n=3)$-loop order. At the
four-loop level, $\gamma_{4\ell}(\alpha_{IR,4\ell})$ tends to be smaller than
$\gamma_{3\ell}(\alpha_{IR,3\ell})$ for values of $N_f$ from $N_{f,b2z}$ to
values of $N_f$ slightly above the middle of the range (\ref{nfirrange}), while
for values of $N_f$ in the upper end of this range,
$\gamma_{4\ell}(\alpha_{IR,4\ell})$ is slightly larger than
$\gamma_{3\ell}(\alpha_{IR,3\ell})$. In general, for the values of $N_f$ where
$\alpha_{IR}$ is sufficiently small that the calculation may be trustworthy,
the value of the anomalous dimension evaluated at the IR zero of the beta
function (both calculated to $n$-loop order)
$\gamma_{n\ell}(\alpha_{IR,n\ell})$, is somewhat smaller than unity.

Several recent high-statistics lattice simulations have been carried out on an
SU(3) gauge theory with a varying number $N_f$ of fermions in the fundamental
representation in the range $6 \le N_f \le 12$ \cite{afn}-\cite{kuti},
\cite{ascalc}, \cite{lgtrev}.  This work has yielded evidence for a regime of
slowly running gauge couplings for $N_f \lsim 12$, consistent with the
presence of an IR zero of the beta function, in agreement with the earlier
continuum estimates in Ref. \cite{chipt}.  Ref. \cite{afn} also found a
considerable enhancement of $\langle \bar\psi\psi\rangle/f_P^3$ in the SU(3)
theory with $N_f=6$.  Further lattice simulations and analysis of data should
yield values of $\gamma_m$ that can be compared with our higher-loop
calculations in this paper.  A preliminary study of the SU(2) theory with
$N_f=6$ fermions has also been reported \cite{deldebbio}.

\section{Adjoint Representation}

In this section we analyze the SU($N$) theory with $N_f$ copies of a Dirac
fermion, or equivalently, $2N_f$ copies of a Majorana fermion, in the adjoint
representation.  For this case, the general expression for the maximal value of
$N_f$ allowed by the requirement of asymptotic freedom, Eq. (\ref{nfmax}),
reduces to
\beq
N_{f,max,adj} = \frac{11}{4} \ , 
\label{nfmax_adj}
\eeq
i.e., restricting $N_f$ to the integers, $N_{f,max}=2$.  The general expression
in Eq. (\ref{nfb2z}) for the value of $N_f$ at which $b_2$ changes sign from 
positive to negative with increasing $N_f$ reduces to 
\beq
N_{f,b2z,adj} = \frac{17}{16} = 1.0625 \ . 
\label{nfb2z_adj}
\eeq
Hence there is only one (integer) value of $N_f$, namely $N_f=2$ Dirac fermions
(equivalently, $N_f=4$ Majorana fermions), for which the
theory is asymptotically free and has an IR zero of the two-loop beta function.
This zero occurs at
\beq
\alpha_{IR,2\ell,adj} = \frac{2\pi}{5N} \simeq \frac{1.257}{N} 
\quad {\rm for} \ N_f=2 \ .  
\label{alfir_2loop_adj_nf2}
\eeq

Specializing the general formula for the critical coupling $\alpha_{cr,R}$ from
the one-gluon exchange approximation to the Dyson-Schwinger equation,
Eq. (\ref{alfcrit}) (see Appendix II) for the present case where $R$ is the
adjoint representation, one obtains $\alpha_{cr,adj}=\pi/(3N)$.  Formally
setting $\alpha_{IR,2\ell,adj}=\alpha_{adj,cr}$ yields the corresponding
estimate for the critical number $N_{f,cr}=83/40 = 2.075$.  This may be rounded
off to the nearest integer, giving $N_{f,cr}$ for the adjoint representation.
In view of the theoretical uncertainty in such an estimate, due to the
strong-coupling nature of the physics involved, an SU($N$) gauge theory with
$N_f=2$ adjoint fermions could be either slightly inside the chirally broken,
confined side of $N_{f,cr}$ or slightly on the other side, where the theory is
chirally symmetric and the evolution into the infrared is governed by an exact
conformal IR fixed point.

For the present case of $N_f=2$ fermions in the adjoint representation of
SU($N$), the coefficients of the beta function are $b_1 =N$, $b_2=-10N^2$,
$b_3=-(101/2)N^3$, and
\beqs
b_4 & = & \bigg [ \frac{5977}{54}N^4+\frac{64}{27}N^3-\frac{448}{27}N^2 
+ \frac{128}{9}N \cr\cr
& + & \frac{176}{9}-\frac{176}{3N^2} \bigg ] \cr\cr
    & + & \zeta(3)\bigg [ \frac{124}{9}N^4-\frac{208}{9}N^3+\frac{3232}{9}N^2
\cr\cr & - & \frac{416}{3}N-\frac{128}{3}+\frac{128}{N^2} \bigg ] \ . 
\label{b4adjnf2}
\eeqs

At the three-loop level, the beta function has two zeros away from the origin,
one of which is negative and unphysical, and the other of which is the physical
IR zero,
\beq
\alpha_{IR,3\ell,adj} = \frac{4\pi(-10+\sqrt{302} \ )}{101N} \simeq 
\frac{0.9180}{N} \ .  
\label{alfir_3loop_adj}
\eeq
At the four-loop level, the beta function has three zeros away from the origin,
one of which is negative and unphysical, one of which is the four-loop IR zero,
$\alpha_{IR,4\ell,adj}$, and the third of which, denoted 
$\alpha_{4\ell,u}$, occurs at a larger positive value of $\alpha$ and is not
relevant to our study, since it is not reached by evolution of the coupling
starting at small $\alpha$ for large $\mu$. We list the numerical values of 
these zeros in Table \ref{betazero_adj}.

The coefficients $\bar c_\ell$ in Eq. (\ref{gamma}) for $\gamma$ for this case
are $\bar c_1=3N/(2\pi)$, $\bar c_2= (11N^2)/(8\pi^2)$, and $\bar
c_3=-N^3/(2\pi^3)$, with $\bar c_4$ given by
\beqs
\pi^4 \bar c_4 &=& \bigg [ 
-\frac{1817}{512}N^4+\frac{1}{96}N^3-\frac{5}{192}N^2 \cr\cr
         & - & \frac{7}{96}N + \frac{1}{32}+\frac{1}{4N}-\frac{3}{16N^3} 
\bigg ] 
\cr\cr
& + & \zeta(3)\bigg [ \frac{155}{128}N^4 - \frac{5}{64}N^3 + \frac{25}{128}N^2 
\cr\cr
& + & \frac{35}{64}N - \frac{15}{64} - \frac{15}{8N} + \frac{45}{32N^3} 
\bigg ] \ . 
\label{c4adjnf2}
\eeqs
(The term in $\bar c_3$ proportional to $\zeta(3)$ and the terms in $\bar c_4$
proportional to $\zeta(4)$ and $\zeta(5)$ vanish for the adjoint representation
for arbitrary $N_f$.)

Evaluating the two-loop expression in Eq. (\ref{gamma_irfp_2loop}) for
$\gamma_m$ at the IR zero of the beta function, also calculated
at the two-loop level, $\alpha_{IR,2\ell,adj}$, we obtain
\beq
\gamma_{2\ell,adj}(\alpha_{IR,2\ell,adj}) = \frac{(11-4N_f)(277-34N_f+40N_f^2)}
{6(-17+16N_f)^2} \ , 
\label{gamma_irfp_2loop_adj}
\eeq
so that for the $N_f=2$ case of interest here, 
\beq
\gamma_{2\ell,adj}(\alpha_{IR,2\ell,adj}) = \frac{41}{50} = 0.820 \quad 
{\rm for} \ \ N_f=2 \ . 
\label{gamma_irfp_2loop_adj_nf2}
\eeq

It is also of interest to evaluate the two-loop $\gamma_m$ at the value of
$\alpha_{cr}$ from the one-gluon exchange (ladder) approximation to the
Dyson-Schwinger equation.  Specializing the general result in
Eq. (\ref{gamma_2loop_at_alfcrit}) to the adjoint representation with $N_f=2$
yields
\beq
\gamma_{2\ell}(\alpha_{cr,adj}) = \frac{47}{72} \simeq 0.653 \ . 
\label{gamma_2loop_at_alfcrit_adj}
\eeq

Evaluating the three-loop result for $\gamma_m$ at 
the IR zero of the beta function calculated at the three-loop level,
$\alpha_{IR,3\ell,adj}$, for the $N_f=2$ case of interest, we obtain 
\beq
\gamma_{3\ell,adj}(\alpha_{IR,3\ell,adj}) = 0.543 \quad 
{\rm for} \ \ N_f=2 \ . 
\label{gamma_irfp_3loop_adj_nf2}
\eeq
which is again independent of $N$.  At the four-loop level, the value of
$\gamma_{4\ell,adj}(\alpha_{IR,4\ell,adj})$ does depend slightly on $N$. We
list the values of these anomalous dimensions in Table \ref{gammavalues_adj}.
The most recent simulations of a lattice gauge theory with SU(2) gauge group
and $N_f=2$ fermions in the adjoint representation report $\gamma_m = 0.49 \pm
0.13$ \cite{cdgk}.  This is in agreement with the calculations of $\gamma_m$
here at the three- and four-loop level, to within the uncertainties of the
respective calculations.

\section{Symmetric and Antisymmetric Rank-2 Tensor Representations}

In this section we consider the SU($N$) theory with $N_f$ fermions in the
symmetric or antisymmetric rank-2 representation, denoted S2 and A2.  Since a
number of formulas are similar for these two cases, we will often give these in
a unified way for both cases, denoted T2 (for rank-2 tensor representation),
with $\pm$ signs distinguishing them.  For S2, our analysis applies for any
$N$, while for A2, we restrict to $N \ge 4$, since the A2 representation is the
singlet for SU(2) and is equivalent to the conjugate fundamental representation
for SU(3).  Note that for SU(4), the A2 representation is self-conjugate. Also,
since for SU(2) the S2 representation is the same as the adjoint
representation, which has already been analyzed, we only consider the
illustrative values $N=3, \ 4$. 

For the two T2 cases, the general expression for the maximal value of
$N_f$ allowed by the requirement of asymptotic freedom, Eq. (\ref{nfmax}),
reduces to
\beq
N_{f,max,T2} = \frac{11N}{2(N \pm 2)} \ , 
\label{nfmax_t2}
\eeq
where the $\pm$ refers to S2 and A2, respectively. As $N$ increases from 2 to
$\infty$, $N_{f,max,S2}$ increases monotonically from 2.75 to $11/2 = 4.5$, and
as $N$ increases from 3 to $\infty$, $N_{f,max,A2}$ decreases monotonically
from 16.5 to the same limit, 4.5. The physical values of $N_{f,max}$ in both
cases are the greatest integral parts of these rational numbers.

For these representations, the general expression in Eq. (\ref{nfb2z}) for the
value of $N_f$ at which the beta function coefficient $b_2$ changes sign from
positive to negative with increasing $N_f$ takes the form
\beq
N_{f,b2z,T2} = \frac{17N^2}{(N \pm 2)(8N \pm 3 - 6N^{-1})} \ . 
\label{nfb2z_t2}
\eeq
As a consequence of the general inequality (\ref{nfdif}), it follows that 
$N_{f,b2z,S2} < N_{f,max,S2}$ and $N_{f,b2z,A2} < N_{f,max,A2}$.  For $N=2$,
the S2 representation is just the adjoint representation, so we only consider
the illustrative values $N=3, \ 4$.  The respective intervals
$N_{f,b2z,S2} < N_f < N_{f,max,S2}$ for which the SU($N$) gauge theory is 
asymptotically free and has an IR zero of $\beta$ are $1.06 < N_f < 2.75$ for
$N=3$ and $1.22 < N_f < 3.30$ for $N=4$. These ranges imply that the only
physical integral values of $N_f$ satisfying these conditions are $N_f=2, \ 3$
for both SU(3) and SU(4).  

For large $N$, $N_{f,b2z,T2}$ has the series expansion
\beq
N_{f,b2z,T2} = \frac{17}{2^3} \mp \frac{323}{2^6N} + \frac{6137}{2^9N^2} 
\mp \frac{103547}{2^{12}N^3} + O \Big ( \frac{1}{N^4} \Big ) 
\label{nfb2z_t2_taylor}
\eeq
As $N$ increases from 2 to $\infty$, $N_{f,b2z,S2}$ increases monotonically
from $17/16=1.0625$ to $17/8=2.125$, and as $N$ increases from 3 to $\infty$,
$N_{f,b2z,A2}$ decreases monotonically from 8.05 to the same limit, 2.125.
This limit is twice the ($N$-independent) value of $N_{f,b2z,adj}=17/16$ for 
the adjoint representation.  Thus, for large $N$, the range (\ref{nfirrange})
where the SU($N$) theory with $N_f$ fermions in the S2 or A2 representation is
asymptotically the same for both, namely, $17/8 < N_f < 11/2$; 
restricting $N_f$ to physical, integer values, this range consists of the 
three values $N_f=3, \ 4, \ 5$

For our further discussion we assume that $N_f$ is in the range $N_{f,b2z,T2} <
N_f < N_{f,max,T2}$ where the theory is asymptotically free and the two-loop
beta function has an IR zero, for the respective cases S2 and A2.  This zero
occurs at the value
\beq
\alpha_{IR,2\ell,T2} = \frac{2\pi(11N-2N_f(N \pm 2)}{-17N^2+N_f(8N^2\pm
  19N \mp 12N^{-1})} \ . 
\label{alfir_2loop_t2}
\eeq

Evaluating Eq. (\ref{alfcrit}) from the ladder approximation to 
the Dyson-Schwinger equation for the fermion propagator for S2 or A2
representations yields the estimate (\ref{alfcrit_t2}) given in Appendix II. 
The resultant $\beta DS$ estimates for $N_{f,cr}$ in the case of the S2
representation and $N=2, \ 3, \ 4$ are $N_{f,cr,S2}=2.1, \ 2.5, \ 2.8$, 
respectively.  For the A2 representation with $N=4$, one has 
$N_{f,cr,A2}=8.1$.

The two-loop expression for the anomalous dimension, evaluated at
$\alpha=\alpha_{IR,2\ell,T2}$, is 
\begin{widetext}
\beqs
& & \gamma_{2\ell,T2}(\alpha_{IR,2\ell,T2}) = \cr\cr
& & \frac{(N \pm 2)(N \mp 1)\Big [ 11N - 2(N\pm 2)N_f \Big ]\bigg [ 
N(554N^2 \pm 99N - 198) + (-34N^3 \pm 22N^2 \mp 360)N_f + 20N(N \pm 2)^2 N_f^2
\bigg ] }
{12 \Big [ -17N^3 +(N \pm 2)(8N^2 \pm 3N - 6)N_f \Big ]^2 } \ . \cr\cr
& & 
\label{gamma_irfp_2loop_t2}
\eeqs
\end{widetext}
We list values of $\gamma_{2\ell,S2}(\alpha_{IR,2\ell},S2)$ for $N=2,3,4$ in
Table \ref{gammavalues_sym} and values of
$\gamma_{2\ell,A2}(\alpha_{IR,2\ell,A2})$ for $N=4$ in Table
\ref{gammavalues_asym}.

It is also of interest to evaluate the two-loop expression for $\gamma$ at
the estimated $\alpha = \alpha_{cr,T2}$.  This yields
\beq
\gamma_{2\ell,T2}(\alpha_{cr,T2}) = 
\frac{322N^2 \pm 225N - 450 -10N(N \pm 2)N_f}{432(N \pm 2)(N \mp 1)} \ . 
\label{gamma_2loop_at_alfcrit_t2}
\eeq
We list these values in Tables \ref{gammavalues_sym} and
\ref{gammavalues_asym}. .

Evaluating the two-loop anomalous dimensions at the two-loop IR zero
of the beta function, $\gamma_{2\ell,T2}(\alpha_{IR,2\ell,T2})$, for $N_f$
equal to the respective BDS-estimated critical values in
Eq. (\ref{nfcr_t2}), we obtain (again with T2 and the $\pm$ signs referring 
respectively to S2 and A2) 
\beqs
& & \gamma_{2\ell,T2}(\alpha_{IR,2\ell,T2})|_{N_f = N_{f,cr,T2}} = \cr\cr
& & \frac{374N^4 \pm 578N^3-931N^2\mp 900N + 900}
 {144(N \pm 2)(N \mp 1)(4N^2 \pm 3N-6)} \ . 
\label{gamma_irfp_2loop_t2_nfcr}
\eeqs
This has the large-$N$ expansion 
\beqs
& & \gamma_{2\ell,T2}(\alpha_{IR,2\ell,T2})|_{N_f = N_{f,cr,T2}} = \cr\cr
& & \frac{187}{288} \mp \frac{17}{128N} + O \bigg ( \frac{1}{N^2} \bigg ) \ . 
\label{gamma_irfp_2loop_t2_nfcr_largeN}
\eeqs
The leading term has the value $187/288 \simeq 0.649$.  

From a lattice study of SU(3) gauge theory with $N_f=2$ fermions in the S2
(sextet) representation, Ref. \cite{dss10} found that this theory is
characterized by slow running behavior consistent with an (exact or
approximate) IR fixed point, and further reported that $\gamma_m < 0.6$ where
it was measured.  For SU(3), the estimate of $\alpha_{cr,S2}$ in
Eq. ({\ref{alfcrit}) gives $\alpha_{cr,S2} = \pi/10 = 0.31$.  Our results for
the IR zero of $\beta$ and the value of $\gamma_m$ at this zero for $N=3$ and
$N_f=2$ are listed in Tables \ref{betazero_sym}. and \ref{gammavalues_sym}. We
find that $\alpha_{IR,n\ell,S2}$ is approximately 0.84 at $n=2$ loop level,
decreases somewhat to 0.50 at three-loop level, and then increases slightly to
0.52 at four-loop level.  The two-loop result for $\gamma_m$ is unphysically
large, while the three- and four-loop values of $\gamma_m$ at the corresponding
three- and four-loop IR zeros of $\beta$ are about 1.3 and 1.4.  These are
somewhat larger than the values reported in Ref. \cite{dss10}, although in
assessing this comparison, one must take account of the significant
strong-coupling uncertainties in our calculation stemming from the fact that
$\alpha_{IR,S2} \sim O(1)$. Our evaluation of the two-loop expression for
$\gamma_m$ at the ladder-Dyson-Schwinger estimate of $\alpha_{cr,S2}$, is 0.65.

\section{Effects of Nonzero Fermion Masses}

The global chiral symmetry that is operative if the fermions are massless, and
the way that it is broken by fermion condensates, is well-known, and we do not
review it here.  However, it is worthwhile to comment on the situation in which
some fermion masses are nonzero.  In this paper we generally assume that the
fermions have zero intrinsic masses in the Lagrangian describing the high-scale
physics, and the only masses that they acquire arise dynamically if they are
involved in condensates that form as the gauge interaction becomes sufficiently
strongly coupled in the infrared.  This is a well-motivated assumption if the
vectorial gauge theory arises as a low-energy effective field theory from an
ultraviolet completion which is a chiral gauge theory.  In turn, this is
natural if the latter theory becomes strongly coupled, since it can then form
fermion condensates that self-break it down to the vectorial subgroup symmetry.
However, one may also choose to focus on the vectorial gauge theory as an
ultraviolet-complete theory in itself. In a vectorial gauge theory, an
intrinsic (bare) mass term for a fermion $\psi$, ${\cal L}_m = -
m\bar\psi\psi$, is allowed by the gauge invariance.  Hence, one may consider a
more general situation in which the fermions may have such intrinsic (hard)
masses in the high-scale Lagrangian \cite{hard}.  In this case, as the
reference scale $\mu$ decreases below the value of the hard mass of some
fermion $m_f$, the beta function changes from one that includes this to one
that excludes this fermion.  If the hard fermion masses are small compared with
the scale $\Lambda$ in the situation where the theory confines and breaks
chiral symmetry spontaneously, then these hard masses have only a small effect.
However, if some of the hard fermion masses are sufficiently large, then as
$\mu$ decreases below their scale and the corresponding fermions are integrated
out of the low-energy theory below this scale, this can significantly change
the infrared properties of the resultant theory.

In applications of slowly running gauge theories to technicolor theories, at
the scale $\Lambda_{TC}$ where the SU($N_{TC}$) gauge coupling grows to O(1)
and is influenced by the presence of an approximate IR zero of the TC beta
function, there can also be non-negligible effects due to four-fermion
operators arising from the higher-lying extended technicolor dynamics
\cite{hillsimmons}-\cite{sanrev}, \cite{etc}-\cite{gnjl,4f}, and these can
affect the scaling properties of $\bar\psi\psi$.  Similar comments apply for
topcolor-assisted technicolor \cite{hillsimmons,tc2uv}.

\section{Conclusions}

In this paper we have studied the evolution of an asymptotically free vectorial
SU($N$) gauge theory from high scales to the infrared taking account of
higher-loop corrections to the beta function and the anomalous dimension
$\gamma_m$ for fermions in the fundamental, adjoint, and rank-2 symmetric and
antisymmetric representations S2 and A2.  We have compared our results with
lower-order calculations.  We have shown that, for fixed $N$ and $N_f$, in the
range for which the two-loop beta function has an IR zero, the value of this
zero decreases as one goes from the two-loop to the three-loop calculations,
and we have determined this decrease quantitatively.  Going further, we have
shown that there is a smaller fractional increase in the value of this IR zero
when calculated to four-loop accuracy, with the final four-loop result still
smaller than the two-loop value. We have analyzed instanton effects and have
demonstrated that they tend to increase the value of the IR zero of the beta
function somewhat.  A major part of our work has been the evaluation of the
anomalous dimension $\gamma_m$ of $\bar\psi\psi$ at the IR zero of the beta
function at the $\ell=2,3,4$ loop levels.  This zero is approximate or exact,
depending on whether for a given $N$, the value of $N_f$ is below or above the
critical value $N_{f,cr}$ below which there is spontaneous chiral symmetry
breaking associated with the formation of a fermion condensate.  We have found
that this $\gamma_m$ at the (approximate or exact) IR zero of the beta function
decreases as one goes from two-loop to three-loop order, and that the four-loop
values also tend to be somewhat less than those at the two-loop level.  The
values that we have calculated for $\gamma_m$ at the IR zero of the beta
function tend to be somewhat smaller than unity.  We have compared our
higher-loop calculations with results from recent lattice simulations and have
found general agreement.  We believe that the higher-loop calculations reported
here should provide a useful reference for comparison with ongoing and future
lattice measurements.

This research was partially supported by the grant NSF-PHY-06-53342.  After
this work was completed, we received the related Ref. \cite{rs2}.  We thank
F. Sannino for sending us this preprint. 

\section{Appendix I} 

For the reader's convenience, we list the three-loop beta
function coefficient, in the $\overline{MS}$ scheme \cite{b3}, 
\begin{widetext}
\beqs
& & b_3 = \frac{2857}{54}C_A^3 +  
+ T_f N_f \bigg[ 2C_f^2 - \frac{205}{9} C_AC_f - \frac{1415}{27}C_A^2 \bigg ]
 + (T_f N_f)^2 \bigg [ \frac{44}{9}C_f + \frac{158}{27}C_A \bigg ] \ . 
\cr\cr
& & 
\label{b3}
\eeqs
The four-loop coefficient is given in Ref. \cite{b4} and is a cubic polynomial
in $N_f$.  We note that the coefficients of the $N_f^0$ (which is independent
of the fermion representation) is positive, and the coefficient of the 
$N_f^3$ term is positive for an arbitrary fermion representation. 

Our normalizations for the quadratic Casimir and trace invariants of a Lie
group are standard.  The quadratic Casimir invariant $C_2(R)$ for the
representation $R$ is given by $\sum_{a=1}^{o(G)} \sum_{j=1}^{dim(R)}
[D_R(T_a)]_{ij} [D_R(T_a)]_{jk} = C_2(R)\delta_{ik}$, where $a,b$ are group
indices, $o(G)$ is the order of the group, $T_a$ are the generators of the
associated Lie algebra, and $D_R(T_a)$ is the matrix form of the $T_a$ in the
representation $R$.  The trace invariant $T(R)$ is defined by
$\sum_{i,j=1}^{dim(R)}[D_R(T_a)]_{ij} [D_R(T_b)]_{ji} = T(R)\delta_{ab}$.

From the calculations of the coefficients of the perturbative expansion of the
anomalous dimension $\gamma_m$ in the $\overline{MS}$ scheme to 
four-loop order in Ref. \cite{gamma4}, we record the three-loop coefficient 
\beq
c_3 =2 C_f \bigg [ \frac{129}{2}C_f^2 - \frac{129}{4}C_fC_A + 
\frac{11413}{108} C_A^2 + C_fT_f N_f (-46+48\zeta(3)) 
 -C_AT_fN_f(\frac{556}{27}+48\zeta(3)) - \frac{140}{27} T_f^2N_f^2 \bigg ] 
\label{c3}
\eeq
We have used the four-loop coefficient $c_4$ from Ref. \cite{gamma4} for our
calculations, but it is too lengthy to reproduce here. 
\end{widetext}

\subsection{Appendix II: Beta-Dyson-Schwinger Estimate of $N_{f,cr}$}

In this appendix we collect some useful known formulas from the one-gluon
exchange (also called ladder) approximation to the Dyson-Schwinger equation for
the fermion propagator \cite{lane,chipt,alm}.  With this approximation, 
a solution of the Dyson-Schwinger in the case of an initial zero input mass of
a fermion in the representation $R$ of the gauge group, yields 
dynamically generated mass if the coupling $\alpha(\mu)$ exceeds a critical
value $\alpha_{cr,R}$ given by \cite{lane,chipt,alm} 
\beq
\alpha_{cr,R} = \frac{\pi}{3 C_f} \ . 
\label{alfcrit}
\eeq
For the representations of interes here, this has the form 
\beq
\alpha_{cr,fund} = \frac{2\pi N}{3(N^2-1)} 
\label{alfcrit_fund}
\eeq
\beq
\alpha_{cr,adj} = \frac{\pi}{3N} 
\label{alfcrit_adj}
\eeq
\beq
\alpha_{cr,T2} = \frac{\pi N}{3(N \pm 2)(N \mp 1)} 
\label{alfcrit_t2}
\eeq
where, as in the text, the $\pm$ signs apply for the S2 and A2 representations,
respectively.

For the illustrative values $N=2, \ 3, \ 4$, this has the following values:
(i) 1.40, 0.79, 0.56 for the fundamental representation; (ii) 
Setting this equal to the two-loop expression for the IR zero of $\beta$ then
yields an estimate for $N_{f,cr}$ to this order, namely 
\beq
N_{f,cr} = \frac{C_A(66C_f+17C_A)}{10T_f(C_A+3C_f)} \ . 
\label{nfcr}
\eeq
For the various representations of interest here, this has the explicit forms
\beq
N_{f,cr,fund} = \frac{2N(50N^2-33)}{5(5N^2-3)}
\label{nfcr_fund}
\eeq
\beq
N_{f,cr,adj} = \frac{83}{40} = 2.075 
\label{nfcr_adj}
\eeq
and
\beq
N_{f,cr,T2} = \frac{N(83N^2 \pm 66N-132)}{5(N \pm 2)(4N^2 \pm 3N -6)}
\label{nfcr_t2}
\eeq
where, as in the text, T2 refers to S2 or A2 and $\pm$ corresponds to S2 and
A2, respectively.  In the same ladder approximation, one finds
$\gamma_m = 1$ at $\alpha = \alpha_{cr,R}$ \cite{chipt} (which
also holds for the DS analysis at a UV-stable fixed point \cite{yam}).
For the gauge group SU($N$) with the illustrative values of $N$ used for the
tables, namely $N=2, \ 3, \ 4$, $N_{f,cr,fund}$ is equal to 7.9, 11.9, and
15.9, respectively, with the large-$N$ form $N_f \sim 4N$.  For S2, the
symmetric rank-2 tensor representation, $N=2, \ 3, \ 4$, $N_{f,cr,S2}$ is equal
to 2.075, 2.5, and 2.9, increasing toward the limit 11/2 = 5.5 in the large-$N$
limit.  In the case of A2, the antisymmetric rank-2 tensor reprepresentation,
for $N=3$, the result is the same as for the fundamental representation, while
for $N=4$, one has $N_{f,crit,A2} \simeq 8.1$, and as $N \to \infty$,
$N_{f,crit,A2}$ decreases toward the limit 11/2.

One understands that, {\it a priori}. there could be significant uncertainty in
these estimates because of the strong-coupling nature of the physics involved
and the one-gluon approximation used for the solution of the Dyson-Schwinger
equation.  Moreover, the DS equation analysis is semi-perturbative in the sense
that it contains polynomial dependence on $\alpha$, and it neglects
nonperturbative effects associated with confinement and instantons.  However,
corrections to the one-gluon exchange approximation have been analyzed and
found not to be too large \cite{alm}.  Recent lattice simulations for SU(3) are
in broad agreement, to within the uncertainties, with the above prediction of
$N_{f,cr} \sim 12$ \ \cite{afn}-\cite{kuti}, \cite{lgtrev}.  Some of the
success of the $\beta DS$ prediction for $N_{f,cr}$ may arise from the fact
that two major physical effects that it ignores would shift $N_{f,cr}$ in
opposite directions and hence tend to cancel each other out \cite{lmax}.  The
DS equation is an integral equation, and the standard analysis of this equation
involves an integration over Euclidean loop momentum $k$ from $k=0$ to
$k=\infty$.  If the theory confines, then the lower bound for the Euclidean
loop momentum should actually not be $k=0$, but instead $k = k_{min.} \sim
r_c^{-1}$ where $r_c$ is the spatial confinement scale \cite{lmax}.  The use of
$k=0$ thus overestimates the tendency toward S$\chi$SB.  Instantons enhance
S$\chi$SB \cite{caldi,cdg,asinstanton,sv}, and the neglect of instanton effects
leads to an underestimate of the tendency toward S$\chi$SB. Since these two
neglected aspects of the physics - confinement and instantons - produce errors
that are of opposite sign as regards the tendency for S$\chi$SB, it is
plausible that these errors tend to cancel out, so that the $\beta DS$ method
may be reasonably accurate in predicting $N_{f,cr}$.

\section{Appendix III - Pad\'e Results}

In this appendix we collect some relevant results on Pad\'e approximants. Given
a Taylor (or asymptotic) series expansion around $z=0$ for the
function $f(z)$,
\beq
f(z) = \sum_{n=0}^{n_{max}} f_n z^n 
\label{fseries}
\eeq
one can construct a set of $[p,q]$ Pad\'e approximants, namely rational 
functions comprised of a numerator polynomial of degree $p$ and a denominator
polynomial of degree $q$, such that $p+q=n_{max}-1$, of the form 
$(\sum_{j=0}^p p_j z^j)/(\sum_{k=0}^q q_k z^k)$.  
Without loss of generality, one can divide numerator and denominator by $q_0$,
so that, after redefinition of the coefficients, one has 
\beq
[p,q]_f(z) = \frac{\sum_{j=0}^p p_j z^j}
                  {1+\sum_{k=1}^q q_k z^k} \ . 
\label{pqpade}
\eeq
The $p+q+1$ coefficients $p_j$ with $0 \le j \le p$ and $q_k$ with $1 \le k \le
q$ are uniquely determined in terms of the $f_n$ coefficients with $0 \le n \le
n_{max}$ by expanding the $[p,q]$ Pad\'e approximant in a Taylor series 
around $z=0$ and solving the set of $n_{max}$ linear equations.

\newpage

\begin{table}
\caption{\footnotesize{Values of the $\ell$-loop beta function coefficients 
$\bar b_\ell$ defined in Eq. (\ref{beta}) in the SU($N$)
gauge theory with $N_f$ fermions transforming according to the fundamental
representation, as functions of $N$ and $N_f$, for the range (\ref{nfrange})
where the theory is asymptotically free.}}
\begin{center}
\begin{tabular}{|c|c|c|c|c|c|} \hline\hline
$N$ & $N_f$& $\bar b_1$ & $\bar b_2$ & $\bar b_3$ & $\bar b_4$ \\ \hline
 2  &  0  & 0.584   & 0.287    & 0.213     & 0.268    \\
 2  &  1  & 0.5305  & 0.235    & 0.154     & 0.191    \\
 2  &  2  & 0.477   & 0.184    & 0.099     & 0.127    \\
 2  &  3  & 0.424   & 0.132    & 0.047     & 0.078    \\
 2  &  4  & 0.371   & 0.080    & $-0.0003$ & 0.044  \\
 2  &  5  & 0.318   & 0.0285   & $-0.044$  & 0.024  \\
 2  &  6  & 0.265   & $-0.023$ & $-0.084$  & 0.020  \\
 2  &  7  & 0.212   & $-0.075$ & $-0.120$  & 0.030 \\
 2  &  8  & 0.159   & $-0.127$ & $-0.152$  & 0.057 \\
 2  &  9  & 0.106   & $-0.178$ & $-0.180$  & 0.099 \\
 2  & 10  & 0.053   & $-0.230$ & $-0.205$  & 0.156 \\
 \hline
 3  &  0  &  0.875  & 0.646    & 0.720     & 1.173    \\
 3  &  1  &  0.822  & 0.566    & 0.582     & 0.910    \\
 3  &  2  &  0.769  & 0.485    & 0.450     & 0.681    \\
 3  &  3  &  0.716  & 0.405    & 0.324     & 0.485    \\
 3  &  4  &  0.663  & 0.325    & 0.205     & 0.322    \\
 3  &  5  &  0.610  & 0.245    & 0.091     & 0.194    \\
 3  &  6  &  0.557  & 0.165    & $-0.016$  & 0.099    \\
 3  &  7  &  0.504  & 0.084    & $-0.118$  & 0.039    \\
 3  &  8  &  0.451  & 0.004    & $-0.213$  & 0.015    \\
 3  &  9  &  0.398  & $-0.076$ & $-0.303$  & 0.025    \\
 3  & 10  &  0.345  & $-0.156$ & $-0.386$  & 0.072    \\
 3  & 11  &  0.292  & $-0.236$ & $-0.463$  & 0.154    \\
 3  & 12  &  0.239  & $-0.317$ & $-0.534$  & 0.273    \\
 3  & 13  &  0.186  & $-0.397$ & $-0.599$  & 0.429    \\
 3  & 14  &  0.133  & $-0.477$ & $-0.658$  & 0.622    \\
 3  & 15  &  0.080  & $-0.557$ & $-0.711$  & 0.852    \\
 3  & 16  &  0.0265 & $-0.637$ & $-0.758$  & 1.121    \\ 
\hline
 4  &  0  &  1.17   & 1.15     & 1.71      & 3.50     \\
 4  &  1  &  1.11   & 1.04     & 1.46      & 2.88     \\
 4  &  2  &  1.06   & 0.932    & 1.22      & 2.31     \\
 4  &  3  &  1.01   & 0.824    & 0.986     & 1.80     \\
 4  &  4  &  0.955  & 0.716    & 0.762     & 1.36     \\
 4  &  5  &  0.902  & 0.607    & 0.546     & 0.972    \\
 4  &  6  &  0.849  & 0.499    & 0.339     & 0.647    \\
 4  &  7  &  0.796  & 0.391    & 0.140     & 0.385    \\
 4  &  8  &  0.743  & 0.283    & $-0.051$  & 0.184    \\
 4  &  9  &  0.690  & 0.175    & $-0.234$  & 0.046    \\
 4  & 10  &  0.637  & 0.066    & $-0.409$  & $-0.029$ \\
 4  & 11  &  0.584  & $-0.042$ & $-0.575$  & $-0.040$ \\
 4  & 12  &  0.531  & $-0.150$ & $-0.733$  & 0.013    \\
 4  & 13  &  0.477  & $-0.258$ & $-0.883$  & 0.131    \\
 4  & 14  &  0.424  & $-0.366$ & $-1.025$  & 0.314    \\
 4  & 15  &  0.371  & $-0.474$ & $-1.16$   & 0.562    \\
 4  & 16  &  0.318  & $-0.583$ & $-1.28$   & 0.877    \\
 4  & 17  &  0.265  & $-0.691$ & $-1.40$   & 1.26     \\
 4  & 18  &  0.212  & $-0.799$ & $-1.51$   & 1.71     \\
 4  & 19  &  0.159  & $-0.907$ & $-1.61$   & 2.22     \\
 4  & 20  &  0.106  & $-1.015$ & $-1.70$   & 2.81     \\
 4  & 21  &  0.053  & $-1.124$ & $-1.79$   & 3.46     \\ 
\hline\hline
\end{tabular}
\end{center}
\label{bvalues}
\end{table}

\begin{table}
\caption{\footnotesize{Values of $N_{f,b2z}$, $N_{f,b3z,\pm}$, and 
$N_{f,b4z,j}$, $i=2,3$, for SU($N$) with 
$N_f$ fermions in the fundamental representation. We only list physical, i.e.,
real, non-negative values. Thus, since $N_{f,bz4,1} < 0$, is is not included.}}
\begin{center}
\begin{tabular}{|c|c|c|c|c|} \hline\hline
$N$ & $N_{f,max}$ & $N_{f,b2z}$ & $(N_{f,b3z,-},N_{f,b3z,+})$ & 
$(N_{f,b4z,2},N_{f,b4z,3})$ \\ \hline
 2  & 11   & 5.55  & (3.99, \ 27.6)  &  none           \\
 3  & 16.5 & 8.05  & (5.84, \ 40.6)  &  none           \\
 4  & 22   & 10.61 & (7.73, \ 53.8)  &  (9.51,11.83)   \\
\hline\hline
\end{tabular}
\end{center}
\label{nfbz_fund}
\end{table}

\begin{table}
\caption{\footnotesize{Estimates of $\alpha_{cr,R}$ from the one-gluon exchange
approximation to the Dyson-Schwinger equation for the fermion propagator.
Values are listed for SU($N$) with $2 \le N \le 6$ and the representations $R=$
(i) fundamental (fund), (ii) adjoint (adj), (iii) symmetric rank-2 tensor (S2),
and (iv) antisymmetric rank-2 (A2).}}
\begin{center}
\begin{tabular}{|c|c|c|c|c|} \hline\hline
$N$ & $\alpha_{cr,fund}$ & $\alpha_{cr,adj}$ & $\alpha_{cr,S2}$ & 
$\alpha_{cr,A2}$ \\ \hline
 2  & 1.40  & 0.52  & 0.52   &  $-$   \\
 3  & 0.79  & 0.35  & 0.31   &  0.79  \\
 4  & 0.56  & 0.26  & 0.23   &  0.42  \\
 5  & 0.44  & 0.21  & 0.19   &  0.29  \\
 6  & 0.36  & 0.17  & 0.16   &  0.22  \\
\hline\hline
\end{tabular}
\end{center}
\label{alfcrittable}
\end{table}

\begin{table}
\caption{\footnotesize{Values of the (approximate or exact) IR zeros in
$\alpha$ of the SU($N$) beta function with $N_f$ fermions in the fundamental
representation, for $N=2,3,4$, calculated at $n$-loop order, and denoted as
$\alpha_{IR,n\ell}$.  For each $N$, we only give results for the integral $N_f$
values in the range (\ref{nfirrange}) where the theory is asymptotically free
and the two-loop beta function has an infrared zero.  For the four-loop beta
function, the cubic equation (\ref{beta_zero_4loop}) has three zeros, one of
which is negative, one of which is $\alpha_{IR,4\ell}$, and the third of which
is positive but farther from the origin.  We include the latter, denoted as
$\alpha_{4\ell,u}$.  We also list zeros from the [1,2] and
[2,1] Pad\'e approximants to the four-loop beta function.}}
\begin{center}
\begin{tabular}{|c|c|c|c|c|c|c|c|} \hline\hline
$N$ & $N_f$& $\alpha_{IR,2\ell}$ & $\alpha_{IR,3\ell}$ &$\alpha_{IR,4\ell}$ 
& $\alpha_{IR,4\ell,[1,2]}$ &  $\alpha_{IR,4\ell,[2,1]}$ &  
$\alpha_{4\ell,u}$ \\ \hline
 2  &  7  &  2.83   & 1.05   & 1.21   & 2.30  & 1.16   & 4.12   \\
 2  &  8  &  1.26   & 0.688  & 0.760  & 0.952 & 0.741  & 3.11   \\
 2  &  9  &  0.595  & 0.418  & 0.444  & 0.475 & 0.438  & 2.395  \\
 2  & 10  &  0.231  & 0.196  & 0.200  & 0.202 & 0.200  & 1.97   \\
 \hline
 3  & 10  &  2.21   & 0.764  & 0.815  & 1.47  & 0.807  & 5.62   \\
 3  & 11  &  1.23   & 0.578  & 0.626  & 0.871 & 0.616  & 3.29   \\
 3  & 12  &  0.754  & 0.435  & 0.470  & 0.561 & 0.462  & 2.295  \\
 3  & 13  &  0.468  & 0.317  & 0.337  & 0.367 & 0.333  & 1.78   \\
 3  & 14  &  0.278  & 0.215  & 0.224  & 0.231 & 0.222  & 1.48   \\
 3  & 15  &  0.143  & 0.123  & 0.126  & 0.127 & 0.125  & 1.29   \\
 3  & 16  &  0.0416 & 0.0397 & 0.0398 & 0.0398& 0.0398 & 1.15   \\ 
\hline
 4  & 13  &  1.85   & 0.604  & 0.628  & 1.14   & 0.625  & 6.94  \\
 4  & 14  &  1.16   & 0.489  & 0.521  & 0.776  & 0.516  & 3.49  \\
 4  & 15  &  0.783  & 0.397  & 0.428  & 0.556  & 0.422  & 2.30  \\
 4  & 16  &  0.546  & 0.320  & 0.345  & 0.407  & 0.340  & 1.73  \\
 4  & 17  &  0.384  & 0.254  & 0.271  & 0.298  & 0.267  & 1.40  \\
 4  & 18  &  0.266  & 0.194  & 0.205  & 0.215  & 0.203  & 1.19  \\
 4  & 19  &  0.175  & 0.140  & 0.145  & 0.149  & 0.145  & 1.05  \\
 4  & 20  &  0.105  & 0.091  & 0.092  & 0.0930 & 0.0921 & 0.947 \\
 4  & 21  &  0.0472 & 0.044  & 0.044  & 0.0444 & 0.0443 & 0.870 \\ 
\hline\hline
\end{tabular}
\end{center}
\label{betazero}
\end{table}

\begin{table}
\caption{\footnotesize{Values of the $\ell$-loop coefficients $ \bar c_\ell$
in the series expansion (\ref{gamma}) for the anomalous dimension
$\gamma_m$, as functions of $N$ and $N_f$, for the range
(\ref{nfrange}) where the theory is asymptotically free.}}
\begin{center}
\begin{tabular}{|c|c|c|c|c|c|} \hline\hline
$N$ & $N_f$& $\bar c_1$ & $\bar c_2$ & $\bar c_3$ & $\bar c_4$ \\ \hline
 2  &  0  & 0.358   & 0.318  & 0.310    & 0.329  \\
 2  &  1  & 0.358   & 0.302  & 0.254    & 0.234  \\
 2  &  2  & 0.358   & 0.286  & 0.195    & 0.143  \\
 2  &  3  & 0.358   & 0.270  & 0.134    & 0.0577 \\
 2  &  4  & 0.358   & 0.254  & 0.0712   & $-0.0218$  \\
 2  &  5  & 0.358   & 0.239  & 0.00656  & $-0.0952$  \\
 2  &  6  & 0.358   & 0.223  & $-0.0601$ & $-0.162$  \\
 2  &  7  & 0.358   & 0.207  & $-0.129$  & $-0.222$  \\
 2  &  8  & 0.358   & 0.191  & $-0.199$  & $-0.274$   \\
 2  &  9  & 0.358   & 0.175  & $-0.272$  & $-0.319$   \\
 2  & 10  & 0.358   & 0.1595 & $-0.346$  & $-0.355$   \\
 \hline
 3  &  0  &  0.637  & 0.853    & 1.26      & 2.03     \\
 3  &  1  &  0.637  & 0.825    & 1.11      & 1.64     \\
 3  &  2  &  0.637  & 0.796    & 0.957     & 1.27     \\
 3  &  3  &  0.637  & 0.768    & 0.801     & 0.909    \\
 3  &  4  &  0.637  & 0.740    & 0.642     & 0.561    \\
 3  &  5  &  0.637  & 0.712    & 0.479     & 0.227    \\
 3  &  6  &  0.637  & 0.684    & 0.312     & $-0.0926$ \\
 3  &  7  &  0.637  & 0.656    & 0.142     & $-0.396$  \\
 3  &  8  &  0.637  & 0.628    & $-0.0313$ & $-0.683$  \\
 3  &  9  &  0.637  & 0.599    & $-0.208$  & $-0.953$  \\
 3  & 10  &  0.637  & 0.571    & $-0.389$  & $-1.21$   \\
 3  & 11  &  0.637  & 0.543    & $-0.573$  & $-1.44$   \\
 3  & 12  &  0.637  & 0.515    & $-0.760$  & $-1.65$   \\
 3  & 13  &  0.637  & 0.487    & $-0.951$  & $-1.85$   \\
 3  & 14  &  0.637  & 0.459    & $-1.145$  & $-2.02$   \\
 3  & 15  &  0.637  & 0.431    & $-1.34$   & $-2.18$   \\
 3  & 16  &  0.637  & 0.402    & $-1.54$   & $-2.31$   \\ 
\hline
 4  &  0  &  0.895   & 1.60   & 3.17     & 6.86     \\
 4  &  1  &  0.895   & 1.56   & 2.89     & 5.88     \\
 4  &  2  &  0.895   & 1.52   & 2.61     & 4.93     \\
 4  &  3  &  0.895   & 1.48   & 2.32     & 4.00     \\
 4  &  4  &  0.895   & 1.44   & 2.03     & 3.09     \\
 4  &  5  &  0.895   & 1.40   & 1.73     & 2.21     \\
 4  &  6  &  0.895   & 1.36   & 1.43     & 1.36     \\
 4  &  7  &  0.895   & 1.33   & 1.12     & 0.526    \\
 4  &  8  &  0.895   & 1.29   & 0.808    & $-0.275$ \\
 4  &  9  &  0.895   & 1.25   & 0.492    & $-1.05$  \\
 4  & 10  &  0.895   & 1.21   & 0.170    & $-1.79$  \\
 4  & 11  &  0.895   & 1.17   & $-0.157$ & $-2.50$  \\
 4  & 12  &  0.895   & 1.13   & $-0.488$ & $-3.18$  \\
 4  & 13  &  0.895   & 1.09   & $-0.825$ & $-3.83$  \\
 4  & 14  &  0.895   & 1.05   & $-1.17$  & $-4.45$  \\
 4  & 15  &  0.895   & 1.01   & $-1.51$  & $-5.025$  \\
 4  & 16  &  0.895   & 0.969  & $-1.86$  & $-5.57$   \\
 4  & 17  &  0.895   & 0.930  & $-2.22$  & $-6.08$   \\
 4  & 18  &  0.895   & 0.890  & $-2.58$  & $-6.54$  \\
 4  & 19  &  0.895   & 0.850  & $-2.95$  & $-6.97$  \\
 4  & 20  &  0.895   & 0.811  & $-3.32$  & $-7.37$  \\
 4  & 21  &  0.895   & 0.771  & $-3.69$  & $-7.71$  \\ 
\hline\hline
\end{tabular}
\end{center}
\label{cvalues}
\end{table}

\begin{table}
\caption{\footnotesize{Values of the anomalous dimension in the SU($N$) theory
with $N_f$ fermions in the fundamental representation, $\gamma_m$,
calculated to the $n$-loop order in perturbation theory and evaluated at the
IR zero of the beta function calculated to this order, $\alpha_{IR,n\ell}$, for
$\ell=2,3,4$.  We denote these as $\gamma_{n\ell}(\alpha_{IR,n\ell})$.  For 
sufficiently small $N_f > N_{f,b2z}$ in each $N$ case, $\alpha_{IR,2\ell}$ is
so large that the formal value of $\gamma_{2\ell}(\alpha_{IR,2\ell})$ is larger
than 2 and hence unphysical; we indicate this by placing these values in 
parentheses.}}
\begin{center}
\begin{tabular}{|c|c|c|c|c|} \hline\hline
$N$ & $N_f$& $\gamma_{2\ell}(\alpha_{IR,2\ell})$ & 
             $\gamma_{3\ell}(\alpha_{IR,3\ell})$ & 
             $\gamma_{4\ell}(\alpha_{IR,4\ell})$ 
  \\ \hline
 2  &  7  & (2.67) & 0.457  & 0.0325  \\
 2  &  8  & 0.752  & 0.272  & 0.204   \\
 2  &  9  & 0.275  & 0.161  & 0.157   \\
 2  & 10  & 0.0910 & 0.0738 & 0.0748  \\
 \hline
 3  & 10  & (4.19) & 0.647  & 0.156   \\
 3  & 11  & 1.61   & 0.439  & 0.250   \\
 3  & 12  & 0.773  & 0.312  & 0.253   \\
 3  & 13  & 0.404  & 0.220  & 0.210   \\
 3  & 14  & 0.212  & 0.146  & 0.147   \\
 3  & 15  & 0.0997 & 0.0826 & 0.0836  \\
 3  & 16  & 0.0272 & 0.0258 & 0.0259  \\
\hline
 4  & 13  & (5.38) & 0.755  & 0.192   \\
 4  & 14  & (2.45) & 0.552  & 0.259   \\
 4  & 15  & 1.32   & 0.420  & 0.281   \\
 4  & 16  & 0.778  & 0.325  & 0.269   \\
 4  & 17  & 0.481  & 0.251  & 0.234   \\
 4  & 18  & 0.301  & 0.189  & 0.187   \\
 4  & 19  & 0.183  & 0.134  & 0.136   \\  
 4  & 20  & 0.102  & 0.0854 & 0.0865  \\
 4  & 21  & 0.0440 & 0.0407 & 0.0409  \\
\hline\hline
\end{tabular}
\end{center}
\label{gammavalues} 
\end{table}

\begin{table}
\caption{\footnotesize{Values of the (approximate or exact) IR zeros in
$\alpha$ of the SU($N$) beta function with $N_f=2$ fermions in the adjoint
representation, for $N=2,3,4$, calculated at $n$-loop order, and denoted as
$\alpha_{IR,n\ell,adj}$. For the four-loop beta function, the cubic equation
(\ref{beta_zero_4loop}) has three zeros, one of which is negative, one of which
is $\alpha_{IR,4\ell,adj}$, and the third of which is positive but farther from
the origin.  We include the latter, denoted as $\alpha_{4\ell,u,adj}$.  We also
list zeros from the [1,2] and [2,1] Pad\'e approximants to the
four-loop beta function.}}
\begin{center}
\begin{tabular}{|c|c|c|c|c|c|c|} \hline\hline
$N$ & $\alpha_{IR,2\ell,adj}$ & $\alpha_{IR,3\ell,adj}$ &
$\alpha_{IR,4\ell,adj}$ & $\alpha_{IR,4\ell,[1,2],adj}$ &
$\alpha_{IR,4\ell,[2,1],adj}$ &  $\alpha_{4\ell,u,adj}$ \\ \hline
 2  &  0.628  & 0.459   & 0.493  & 0.522  & 0.484   & 2.235  \\
 3  &  0.419  & 0.306   & 0.323  & 0.341  & 0.320   & 1.80   \\
 4  &  0.314  & 0.2295  & 0.241  & 0.253  & 0.239   & 1.47   \\
\hline\hline
\end{tabular}
\end{center}
\label{betazero_adj}
\end{table}

\begin{table}
\caption{\footnotesize{Values of the anomalous dimension $\gamma_m$ in
an SU($N$) gauge theory with $N_f=2$ (Dirac) fermions in the adjoint 
representation, calculated to the $n$-loop
order in perturbation theory and evaluated at the IR zero of the beta function
calculated to this order, for $n=2,3,4$.  We denote
these as $\gamma_{n\ell,adj}(\alpha_{IR,n\ell,adj})$.  We also list 
the value of $\gamma_{2\ell,adj}$ evaluated at $\alpha$ equal to the 
$\beta DS$ estimate, Eq. (\ref{alfcrit}), for $\alpha_{cr,adj})$.}}
\begin{center}
\begin{tabular}{|c|c|c|c|c|} \hline\hline
$N$ & $\gamma_{2\ell,adj}(\alpha_{IR,2\ell,adj})$ & 
      $\gamma_{3\ell,adj}(\alpha_{IR,3\ell,adj})$ & 
      $\gamma_{4\ell,adj}(\alpha_{IR,4\ell,adj})$ &
      $\gamma_{2\ell,adj}(\alpha_{cr,adj})$  \\ 
\hline
 2  &  0.820  & 0.543   & 0.571  &  0.653 \\
 3  &  0.820  & 0.543   & 0.561  &  0.653 \\
 4  &  0.820  & 0.543   & 0.557  &  0.653 \\
\hline\hline
\end{tabular}
\end{center}
\label{gammavalues_adj}
\end{table}

\begin{table}
\caption{\footnotesize{Values of the (approximate or exact) IR zero in
$\alpha$ of the SU($N$) beta function with $N_f=2$ fermions in the symmetric
rank-2 (i.e., S2) representation, for $N=3,4$, calculated at $n$-loop order, 
and denoted as $\alpha_{IR,n\ell,S2}$.}}
\begin{center}
\begin{tabular}{|c|c|c|c|c|} \hline\hline
$N$ & $N_f$ & $\alpha_{IR,2\ell,S2}$ & $\alpha_{IR,3\ell,S2}$ & 
$\alpha_{IR,4\ell,S2}$ \\ \hline
 3  &  2  &  0.842   & 0.500   & 0.522  \\
 3  &  3  &  0.085   & 0.079   & 0.080  \\
 4  &  2  &  0.967   & 0.485   & 0.485  \\
 4  &  3  &  0.152   & 0.129   & 0.132  \\
\hline\hline
\end{tabular}
\end{center}
\label{betazero_sym}
\end{table}

\begin{table}
\caption{\footnotesize{Values of the (approximate or exact) IR zero in
$\alpha$ of the SU(4) beta function with $N_f$ fermions in the antisymmetric
rank-2 (i.e., A2) representation, for the range $5 \le N_f \le 10$ where the
theory is asymptotically free and has an IR zero of the beta function, 
calculated at $n$-loop order, and denoted as $\alpha_{IR,n\ell,A2}$.}}
\begin{center}
\begin{tabular}{|c|c|c|c|c|} \hline\hline
$N$ & $N_f$ & $\alpha_{IR,2\ell,A2}$ & $\alpha_{IR,3\ell,A2}$ & 
$\alpha_{IR,4\ell,A2}$  \\ \hline
 4  &  6  &  2.17   & 0.664   & 0.723  \\
 4  &  7  &  0.890  & 0.437   & 0.482  \\
 4  &  8  &  0.449  & 0.287   & 0.312  \\
 4  &  9  &  0.225  & 0.174   & 0.183  \\
 4  & 10  &  0.090  & 0.080   & 0.082  \\
\hline\hline
\end{tabular}
\end{center}
\label{betazero_asym}
\end{table}

\begin{table}
\caption{\footnotesize{Values of $\gamma_m$ in
an SU($N$) gauge theory with $N_f$ fermions in the symmetric rank-2
tensor representation S2, calculated to the $n$-loop
order in perturbation theory and evaluated at the IR zero of the beta function
calculated to this order, for $n=2,3,4$.  We denote
these as $\gamma_{n\ell,S2}(\alpha_{IR,n\ell,S2})$. We also list 
$\gamma_{2\ell,S2}$ evaluated at $\alpha$ equal to the estimate 
Eq. (\ref{alfcrit}) for $\alpha_{cr,S2}$.}}
\begin{center}
\begin{tabular}{|c|c|c|c|c|c|} \hline\hline
$N$ & $N_f$ & $\gamma_{2\ell,S2}(\alpha_{IR,2\ell,S2})$ & 
              $\gamma_{3\ell,S2}(\alpha_{IR,3\ell,S2})$ & 
              $\gamma_{4\ell,S2}(\alpha_{IR,4\ell,S2})$ & 
              $\gamma_{2\ell,S2}(\alpha_{cr,S2})$  \\ \hline
 3  &  2  & (2.44)  & 1.28   & 1.38   &  0.653    \\
 3  &  3  & 0.144   & 0.133  & 0.134  &  0.619    \\
 4  &  2  & (4.82)  & (2.08) & (2.27) &  0.659    \\
 4  &  3  & 0.381   & 0.313  & 0.319  &  0.629    \\
\hline\hline
\end{tabular}
\end{center}
\label{gammavalues_sym}
\end{table}

\begin{table}
\caption{\footnotesize{Values of $\gamma_m$ in an SU($N$) gauge theory with
$N_f$ fermions in the antisymmetric rank-2 tensor representation A2,
calculated to the $n$-loop order in perturbation theory and evaluated at the
IR zero of the beta function calculated to this order, for $N=4$ and
$n=2,3,4$.  We denote these as $\gamma_{n\ell,A2}(\alpha_{IR,n\ell,A2})$. We
also list $\gamma_{2\ell,A2}$ evaluated at $\alpha$ equal to the estimate
Eq. (\ref{alfcrit}) for $\alpha_{cr,A2}$.}}
\begin{center}
\begin{tabular}{|c|c|c|c|c|c|} \hline\hline
$N$ & $N_f$& $\gamma_{2\ell,A2}(\alpha_{IR,2\ell,A2})$ & 
             $\gamma_{3\ell,A2}(\alpha_{IR,3\ell,A2})$ & 
             $\gamma_{4\ell,A2}(\alpha_{IR,4\ell,A2})$ &
             $\gamma_{2\ell,A2}(\alpha_{cr,A2})$ 
  \\ \hline
 4  &  6  & (9.78) & 1.38   & 0.647  & 0.769   \\
 4  &  7  & (2.19) & 0.695  & 0.505  & 0.750   \\
 4  &  8  & 0.802  & 0.402  & 0.375  & 0.732  \\
 4  &  9  & 0.331  & 0.228  & 0.231  & 0.713   \\
 4  & 10  & 0.117  & 0.101  & 0.102  & 0.695   \\
\hline\hline
\end{tabular}
\end{center}
\label{gammavalues_asym}
\end{table}


\begin{thebibliography}{99}

\bibitem{chipt}
T. Appelquist, D. Karabali, and L. C. R. Wijewardhana, Phys. Rev. Lett. {\bf
57}, 957 (1986); T. Appelquist and L. C. R. Wijewardhana, Phys. Rev. D
{\bf 35}, 774 (1987);  Phys.  Rev. D {\bf 36}, 568 (1987); 
T. Appelquist, J. Terning, and L. C. R. Wijewardhana,
Phys. Rev. Lett. {\bf 77}, 1214 (1996). 

\bibitem{alm}
T. Appelquist, K. Lane, and U. Mahanta, Phys. Rev. Lett. {\bf 61}, 1553 
(1988). 

\bibitem{bv}
T. A. Ryttov and R. Shrock, Phys. Rev. D {\bf 81}, 116003 (2010); 
erratum, {\it ibid.} D {\bf 82}, 059903 (2010). 

\bibitem{hillsimmons}
C. T. Hill and E. H. Simmons, Phys. Rep. {\bf 381}, 235 (2003).

\bibitem{dewsb}
{\it Workshop on Dynamical Electroweak Symmetry Breaking},
Southern Denmark Univ. 2008 (http://hep.sdu.dk/dewsb).

\bibitem{sanrev}
F. Sannino, Acta Phys. Polon. B {\bf 40}, 3533 (2009). 

\bibitem{scalc}
T. Appelquist and F. Sannino, Phys. Rev. D {\bf 59}, 067702 (1999);
M. Harada, M. Kurachi, and K. Yamawaki, Phys. Rev. {\bf 70}, 033009 (2004);
M. Kurachi and R. Shrock, Phys. Rev. D {\bf 74}, 056003 (2006).

\bibitem{ascalc}
T. Appelquist et al., arXiv:1009.5967. 

\bibitem{higherrep}
%
K. Lane and E. Eichten, Phys. Lett. B {\bf 222}, 274 (1989); 
D. Hong, S. Hsu, and F. Sannino, Phys. Lett. B {\bf 597} (2004) 89; 
F. Sannino and K. Tuominen, Phys. Rev. D {\bf 71}, 051901(R) (2005); 
N. D. Christensen and R. Shrock, Phys. Lett. B {\bf 632}, 92 (2006); 
D. D. Dietrich and F. Sannino, Phys. Rev. D {\bf 75}, 085018 (2007);
R. Foadi, M. T. Frandsen, T. A. Ryttov, F. Sannino, Phys. Rev.  D {\bf 76}, 
055005 (2007); 
T. A. Ryttov and F. Sannino, Phys. Rev. D {\bf 76}, 105004 (2007).

\bibitem{etc}
%
T. Appelquist and J. Terning, Phys. Rev. D {\bf 50}, 2116 (1994);
T. Appelquist and R. Shrock,  Phys. Lett. B {\bf 548}, 204 (2002);
T. Appelquist and R. Shrock, Phys. Rev. Lett. {\bf 90}, 201801 (2003);
T. Appelquist, M. Piai, and R. Shrock, Phys. Rev. D {\bf 69}, 015002 (2004);
N. C. Christensen and R. Shrock, Phys. Rev. D {\bf 74}, 015004 (2006);
T. A. Ryttov and R. Shrock, Phys. Rev. D {\bf 81}, 115013 (2010).

\bibitem{afn}
%
 T. Appelquist, G. Fleming, and E. Neil, Phys. Rev. Lett. {\bf 100},
171607 (2008);  T. Appelquist, G. Fleming, and E. Neil, Phys. Rev. D {\bf 79},
076010 (2009); T. Appelquist, Prog. Theor. Phys. {\bf 180}, 72 (2009); 
T. Appelquist et al., Phys. Rev. Lett. {\bf 104}, 071601 (2010); 
arXiv:1009.5967.

\bibitem{lombardo}
A. Deuzeman, M. P. Lombardo, and E. Pallante, Phys. Lett. B {\bf 670}, 41 
(2008); arXiv:0810.1719, arXiv:0904.4662.

\bibitem{mawhinney}
X.-Y. Jin and R. Mawhinney, PoS Lattice-2008:059 (2008), arXiv:0812.0413.

\bibitem{hasenfratz}
A. Hasenfratz, Phys. Rev. D {\bf 80}, 034505 (2009). 

\bibitem{kuti}
Z. Fodor, K. Holland, J. Kuti,
D. Nogradi, and C. Schroeder, Phys. Lett. B {\bf 681}, 353 (2009);
arXiv:0911.2463.

\bibitem{cgss}
%
S. Catterall, J. Giedt, F. Sannino, and J. Schneible, JHEP 0811, 009 (2008);
ArXiv:0910.4387.

\bibitem{dss09}
T. DeGrand, Y. Shamir, and B. Svetitsky, Phys. Rev. D {\bf 79}, 034501 (2009).

\bibitem{dss10}
T. DeGrand, Y. Shamir, and B. Svetitsky, Phys. Rev. D {\bf 82}, 054503
(2010). 

\bibitem{kutisym}
Z. Fodor, K. Holland, J. Kuti, D. Nogradi, and C. Schroeder,
Phys. Lett. B {\bf 681}, 353 (2009); JHEP {\bf 2009}, 103 (2009).

\bibitem{tuominen}
A. J. Hietanen, K. Rummukainen, and K. Tuominen, Phys. Rev. D {\bf 80}, 
094504 (2009).

\bibitem{deldebbio}
L. Del Debbio, B. Lucini, A. Patella, C. Pica, and A. Rago,
Phys. Rev. D {\bf 80}, 074507 (2009); 
F. Bursa, L. Del Debbio, L. Keegan, C. Pica, and T. Pickup,
Phys. Rev. D {\bf 81}, 014505 (2010). 

\bibitem{kogutsinclair}
J. B. Kogut and D. K. Sinclair, ArXiv:1002.2988. 

\bibitem{cdgk}
S. Catterall, L. Del Debbio, J. Giedt, and L. Keegan, arXiv:1010.5909.

\bibitem{lgtrev}
G. Fleming, talk at the Int. Conf. on High Energy Physics ICHEP 2010,
Paris, http://www.ichep2010.fr.

\bibitem{b1}
D. J. Gross and F. Wilczek, Phys. Rev. Lett. {\bf 30}, 1343 (1973); 
H. D. Politzer, Phys. Rev. Lett. {\bf 30}, 1346 (1973); G. 't Hooft,
unpublished. 

\bibitem{b2}
W. E. Caswell, Phys. Rev. Lett. {\bf 33}, 244 (1974); 
D. R. T. Jones, Nucl. Phys. B {\bf 75}, 531 (1974). 

\bibitem{gw2}
D. J. Gross and F. Wilczek, Phys. Rev. D {\bf 8}, 3633 (1973); 
D. J. Gross, in R. Balian and J. Zinn-Justin, eds. {\it Methods in Field
  Theory}, Les Houches 1975 (North Holland, Amsterdam, 1976). 

\bibitem{b3}
O. V. Tarasov, A. A. Vladimirov, and A. Yu. Zharkov, Phys. Lett. B {\bf 93},
429 (1980); S. A. Larin and J. A. M. Vermaseren, Phys. Lett. B {\bf 303}, 334
(1993). 

\bibitem{b4}
T. van Ritbergen, J. A. M. Vermaseren, and S. A. Larin, Phys. Lett. B {\bf
 400}, 379 (1997). 

\bibitem{gamma4}
J. A. M. Vermaseren, S. A. Larin, and T. van Ritbergen, Phys. Lett. B {\bf
  405},
 327 (1997). 

\bibitem{rs}
T. A. Ryttov and F. Sannino, Phys. Rev. D {\bf 78}, 065001 (2008). 

\bibitem{btd}
S. J. Brodsky and G. F. de T\'eramond, in M. Harada, M. Tanabashi, and
K. Yamawaki, eds., {\it The Origin of Mass and Strong Coupling Gauge Theories,
  SCGT2006} (World Scientific, Singapore, 2008), p. 31. 

\bibitem{creutz}
M. Creutz, unpublished. 

\bibitem{lmax}
S. Brodsky and R. Shrock, Phys. Lett. B {\bf 666}, 95 (2008).

\bibitem{thooft76}
G. 't Hooft, Phys. Rev. Lett. {\bf 37}, 8 (1976); 
Phys. Rev. D {\bf 14}, 3432 (1976); erratum {\it ibid.} D {\bf 18}, 2199
(1978). 

\bibitem{cdg}
C. G. Callan, R. F. Dashen, and D. J. Gross, Phys. Rev. D 
{\bf 17}, 2717 (1978); Phys. Rev. D {\bf 20}, 3279 (1979). 

\bibitem{caldi}
D. Caldi, Phys. Rev. Lett. {\bf 39}, 121 (1977). 

\bibitem{asinstanton}
T. Appelquist and S. Selipsky, Phys. Lett. B {\bf 400}, 364 (1997). 

\bibitem{sv}
E. Shuryak and M. Velkovsky, Phys. Lett. B {\bf 437}, 398 (1998). 

\bibitem{bz}
T. Banks and A. Zaks, Nucl. Phys. B {\bf 196}, 189 (1982). 

\bibitem{nfintegral}
%
Here and elsewhere, when expressions are given for $N_f$ that formally evaluate
to non-integral real values, it is understood implicitly that one infers
an appropriate integral value of $N_f$ from them, either the greatest integral
part or the nearest integer, depending on the context.

\bibitem{uvfp}
%
Parenthetically, we observe that if (i) the exact beta function of a theory
were to have a zero at a (nonzero, positive) value $\alpha_1$ with
$d\beta/d\alpha > 0$ at $\alpha_1$, and (ii) another zero at a larger value,
$\alpha_2$ with $d\beta/d\alpha < 0$ at $\alpha_2$ with (iii) $\beta > 0$ for
$\alpha_1 < \alpha < \alpha_2$, and if (iv) the initial condition in the deep
ultraviolet is that as $\mu \to \infty$, $\alpha(\mu)$ approaches $\alpha_2$
from below, then as the scale $\mu$ decreases, $\alpha$ would decrease from the
UV fixed point $\alpha_2$ and approach the IR fixed point $\alpha_1$ from above
as $\mu \to 0$.  This type of behavior is not relevant to our theory with the
initial condition on $\alpha$ that we assume in the ultraviolet. 

\bibitem{ungam}
G. Mack, Commun. Math. Phys. {\bf 55}, 1 (1977);
M. Flato and C. Fronsdal, Lett. Math. Phys. {\bf 8}, 159 (1984);
V. K. Dobrev and V. B. Petkova, Phys. Lett. B {\bf 162}, 127 (1985). 

\bibitem{hard}
%
Here a ``hard'' fermion mass is a mass that would remain in the hypothetical
limit in which one turned off the SU($N$) interactions.  If these masses are
dynamically generated, then they are actually soft on the higher scale where
they arise, as discussed in N. D. Christensen and R. Shrock, Phys. Rev. Lett. 
{\bf 94}, 241801 (2005). 

\bibitem{gnjl}
T. Appelquist, M. Soldate, T. Takeuchi, and
L. C. R. Wijewardhana, in G. Domokos and S. Kovesi-Domokos, eds., {\it 
Proc. Johns Hopkins Workshop on Current Problems in Particle Theory} (World
Scientific, Singapore, 1988); K. I. Kondo, H. Mino, and K. Yamawaki,
Phys. Rev. D {\bf 39}, 2430 (1989). 

\bibitem{4f} M. Kurachi, R. Shrock, and K. Yamawaki, Phys. Rev. D {\bf 76},
035003 (2007); H. S. Fukano and F. Sannino, Phys. Rev. D {\bf 82}, 035021
(2010).

\bibitem{tc2uv}
A. Martin and K. Lane, Phys. Rev. D {\bf 71}, 015011 (2005).
F. Braam, M. Flossdorf, R. S. Chivukula, S. Di Chiara, and E. H. Simmons,
Phys. Rev. D {\bf 77}, 055005 (2008);
T. A. Ryttov and R. Shrock, Phys. Rev. D {\bf 82}, 055012 (2010),

\bibitem{yam}
B. Holdom, Phys. Rev. Lett. {\bf 60}, 1233 (1988); 
K. Yamawaki, M. Bando, and K. Matumoto, Phys. Rev. Lett. {\bf
56}, 1335 (1986); V. Miransky and K. Yamawaki, Phys. Rev. D {\bf 55}, 5051 
(1997); Phys. Rev. {\bf 56}, E 3768 (1997).

\bibitem{lane}
K. Lane, Phys. Rev. D {\bf 10}, 2605 (1974). 

\bibitem{rs2}
C. Pica and F. Sannino, arXiv:1011.3832. 

\end{thebibliography}
\end{document}